\documentclass[english,onecolumn,conference]{IEEEtran}
\usepackage[T1]{fontenc}
\usepackage[latin9]{inputenc}
\usepackage{enumitem}
\usepackage{amsmath}
\usepackage{amsthm}
\usepackage{amssymb}
\usepackage{esint}

\makeatletter
\theoremstyle{plain}
\newtheorem{thm}{\protect\theoremname}
\theoremstyle{definition}
\newtheorem{defn}[thm]{\protect\definitionname}
\theoremstyle{plain}
\newtheorem{prop}[thm]{\protect\propositionname}
\theoremstyle{plain}
\newtheorem{lem}[thm]{\protect\lemmaname}
\theoremstyle{plain}
\newtheorem{conjecture}[thm]{\protect\conjecturename}
\theoremstyle{remark}
\newtheorem{claim}[thm]{\protect\claimname}
\newlist{casenv}{enumerate}{4}
\setlist[casenv]{leftmargin=*,align=left,widest={iiii}}
\setlist[casenv,1]{label={{\itshape\ \casename} \arabic*.},ref=\arabic*}
\setlist[casenv,2]{label={{\itshape\ \casename} \roman*.},ref=\roman*}
\setlist[casenv,3]{label={{\itshape\ \casename\ \alph*.}},ref=\alph*}
\setlist[casenv,4]{label={{\itshape\ \casename} \arabic*.},ref=\arabic*}

\usepackage{mathrsfs}
\usepackage{algorithm,algpseudocode}

\usepackage{colortbl}
\definecolor{lightgray}{rgb}{0.9,0.9,0.9}
\definecolor{lightred}{rgb}{1,0.8,0.8}
\definecolor{lightgreen}{rgb}{0.6,1,0.6}
\definecolor{lightyellow}{rgb}{1,1,0.5}
\definecolor{lightgrey}{rgb}{0.8,0.8,0.8}

\allowdisplaybreaks[1]

\makeatother

\usepackage{babel}
\providecommand{\casename}{Case}
\providecommand{\claimname}{Claim}
\providecommand{\conjecturename}{Conjecture}
\providecommand{\definitionname}{Definition}
\providecommand{\lemmaname}{Lemma}
\providecommand{\propositionname}{Proposition}
\providecommand{\theoremname}{Theorem}

\begin{document}
\title{Infinite Divisibility of Information}
\author{Cheuk Ting Li\\
Department of Information Engineering\\
The Chinese University of Hong Kong\\
Email: ctli@ie.cuhk.edu.hk}
\maketitle
\begin{abstract}
We study an information analogue of infinitely divisible probability
distributions, where the i.i.d. sum is replaced by the joint distribution
of an i.i.d. sequence. A random variable $X$ is called informationally
infinitely divisible if, for any $n\ge1$, there exists an i.i.d.
sequence of random variables $Z_{1},\ldots,Z_{n}$ that contains the
same information as $X$, i.e., there exists an injective function
$f$ such that $X=f(Z_{1},\ldots,Z_{n})$. While there does not exist
informationally infinitely divisible discrete random variable, we
show that any discrete random variable $X$ has a bounded multiplicative
gap to infinite divisibility, that is, if we remove the injectivity
requirement on $f$, then there exists i.i.d. $Z_{1},\ldots,Z_{n}$
and $f$ satisfying $X=f(Z_{1},\ldots,Z_{n})$, and the entropy satisfies
$H(X)/n\le H(Z_{1})\le1.59H(X)/n+2.43$. We also study a new class
of discrete probability distributions, called spectral infinitely
divisible distributions, where we can remove the multiplicative gap
$1.59$. Furthermore, we study the case where $X=(Y_{1},\ldots,Y_{m})$
is itself an i.i.d. sequence, $m\ge2$, for which the multiplicative
gap $1.59$ can be replaced by $1+5\sqrt{(\log m)/m}$. This means
that as $m$ increases, $(Y_{1},\ldots,Y_{m})$ becomes closer to
being spectral infinitely divisible in a uniform manner. This can
be regarded as an information analogue of Kolmogorov's uniform theorem.
Applications of our result include independent component analysis,
distributed storage with a secrecy constraint, and distributed random
number generation.
\end{abstract}

\begin{IEEEkeywords}
Infinitely divisible distributions, information spectrum, independent
component analysis, distributed storage, Kolmogorov's uniform theorem.
\end{IEEEkeywords}

\section{Introduction}

This paper seeks to answer the following question: can a piece of
information be divided into arbitrarily many pieces? Is information
infinitely divisible, like space and time? To answer this question,
we first review the concept of infinite divisibility in probability
theory.

In probability theory, it is often of interest to consider the sum
$\sum_{i=1}^{n}Z_{i}$ of $n$ i.i.d. random variables $Z_{1},\ldots,Z_{n}\in\mathbb{R}$.
Some notable limit theorems regarding i.i.d. sums are the law of large
numbers, central limit theorem (with limit distribution being the
Gaussian distribution) and the law of small numbers \cite{chen1975poisson}
(with limit distribution being the Poisson distribution). One may
also ask the reverse question that whether a given probability distribution
can be expressed as such an i.i.d. sum. This is captured by the concept
of infinite divisibility: a probability distribution $P$ is infinitely
divisible if, for any $n\ge1$, there exists $n$ i.i.d. random variables
$Z_{1},\ldots,Z_{n}$ such that $\sum_{i=1}^{n}Z_{i}$ has distribution
$P$. Kolmogorov's uniform theorem \cite{kolmogorov1956two} states
that the distributions of i.i.d. sums $\sum_{i=1}^{m}Y_{i}$ are uniformly
close to being infinitely divisible, in the sense that for any i.i.d.
random variables $Y_{1},\ldots,Y_{m}\in\mathbb{R}$, there exists
$W$ following an infinitely divisible distribution such that $d_{\mathrm{U}}(F_{\sum_{i}Y_{i}},F_{W})\le cm^{-1/5}$,
where we write $F_{\sum_{i}Y_{i}}$ for the cdf of $\sum_{i=1}^{m}Y_{i}$,
$d_{\mathrm{U}}$ denotes the uniform metric $d_{\mathrm{U}}(F_{1},F_{2}):=\sup_{x}|F_{1}(x)-F_{2}(x)|$,
and $c>0$ is a universal constant. See \cite{prokhorov1960uniform,lecam1965distribution,ibragimov1974rate}
for improvements on Kolmogorov's uniform theorem. Arak \cite{arak1982convergence,arak1983improvement}
improved the bound to $cm^{-2/3}$, which is shown to be tight.

In information theory, the ``sum'' of the information of two independent
random variables $Z_{1},Z_{2}$ is captured by the joint random variable
$(Z_{1},Z_{2})$ instead of $Z_{1}+Z_{2}$. Nevertheless, by considering
the self information $\iota_{Z_{i}}(z):=-\log p_{Z_{i}}(z)$, we have
$\iota_{(Z_{1},Z_{2})}(Z_{1},Z_{2})=\iota_{Z_{1}}(Z_{1})+\iota_{Z_{2}}(Z_{2})$,
and hence the information spectrum of $(Z_{1},Z_{2})$ (the distribution
of $\iota_{(Z_{1},Z_{2})}(Z_{1},Z_{2})$) is the independent sum of
the respective information spectra of $Z_{1}$ and $Z_{2}$ (refer
to \cite{han2003information} for discussion on information spectrum).
Due to this connection, many results on i.i.d. sums in probability
theory can be carried over to information theory. For example, the
asymptotic equipartition property \cite{shannon1948mathematical}
(corresponding to the law of large numbers), large deviations analysis
\cite{sanov1957probability,gallager1965simple,csiszar2011information},
and the Gaussian approximation of the information spectrum \cite{han2003information,hayashi2008second}
(corresponding to the central limit theorem) are important tools in
information theory.

On the reverse question that whether a given random variable can be
expressed as an i.i.d. sequence, bits are regarded as the universal
unit of information, and a piece of information is often decomposed
into bits. Two examples are Huffman coding \cite{huffman1952method}
and the Knuth-Yao method for generating random variables \cite{knuth1976complexity},
both of them establishing that, loosely speaking, any discrete random
variable $X$ can be decomposed into roughly $H(X)=\mathbf{E}[\iota_{X}(X)]$
number of bits. A caveat of both results is that they concern the
variable-length setting, where the number of bits is not fixed and
can depend on the value of $X$ (and $H(X)$ is only the approximate
average number of bits). \footnote{We remark that the asymptotic equipartition property implies that
the i.i.d. sequence $X=(Y_{1},\ldots,Y_{m})$ can be decomposed into
a fixed number $H(X)+o(m)$ of bits with low error probability as
$m\to\infty$. In comparison, the results in this paper also apply
when there is only one random variable (i.e., $m=1$), instead of
an i.i.d. sequence.}

In this paper, we consider the information analogue of infinite divisibility,
where the number of components $n$ cannot depend on the value of
$X$. Given a random variable $X$, we would determine whether, for
all $n\ge1$, there exists an i.i.d. sequence of random variables
$Z_{1},\ldots,Z_{n}$ that contains the same information as $X$,
i.e., $H(X|Z_{1},\ldots,Z_{n})=H(Z_{1},\ldots,Z_{n}|X)=0$ (there
exists an injective function $f$ such that $X=f(Z_{1},\ldots,Z_{n})$,
or equivalently, they generate the same $\sigma$-algebra). If this
is possible, then these $Z_{1},\ldots,Z_{n}$ are an equipartition
of the information in $X$, and they can be regarded as ``$X^{1/n}$''
(the inverse operation of taking $n$ i.i.d. copies $X^{n}=(X_{1},\ldots,X_{n})$).

Unfortunately, informationally infinitely divisible discrete random
variable does not exist, revealing a major difference between the
``sum'' of information and the sum of real random variables.\footnote{We remark that any continuous random variable over $\mathbb{R}$ is
informationally infinitely divisible due to the existence of a measure-preserving
bijection between $[0,1]^{n}$ and $[0,1]$ modulo zero.} This is due to the fact that, while any random variable has an information
spectrum, not all distributions are the information spectrum of some
random variables.

Nevertheless, we show that any random variable has a bounded multiplicative
gap to infinite divisibility. More precisely, for any discrete random
variable $X$ and $n\in\mathbb{N}$, there exists i.i.d. discrete
random variables $Z_{1},\ldots,Z_{n}$ such that $H(X|Z_{1},\ldots,Z_{n})=0$,
and
\begin{equation}
\frac{1}{n}H(X)\le H(Z_{1})\le\frac{1}{n}\cdot\frac{e}{e-1}H(X)+2.43,\label{eq:intro_hzbd}
\end{equation}
where the upper bound has a multiplicative gap $e/(e-1)\approx1.582$
from the lower bound. The additive gap $2.43$ can be replaced with
a term that scales like $O(n^{-1/2}\log n)$ as $n\to\infty$ for
any fixed $H(X)$. This is proved in Theorem \ref{thm:pin_gap}. As
a consequence, the random variable $X$ can indeed be divided into
$n$ i.i.d. pieces (where the entropy of each piece tends to $0$
as $n\to\infty$), but at the expense of a multiplicative gap $\approx1.582$
for large $H(X)$, and the multiplicative gap becomes worse (and tends
to $\infty$) for small $H(X)$.

We also study a new class of discrete probability distributions, called
the \emph{spectral infinitely divisible }(SID) distributions, defined
as the class of distributions where the information spectrum has an
infinitely divisible distribution. Examples include the discrete uniform
distribution and the geometric distribution. We show that if $X$
follows an SID distribution, then the multiplicative gap $e/(e-1)$
in \eqref{eq:intro_hzbd} can be eliminated (though the additive gap
$2.43$ stays).

Furthermore, we can study the case where $X=(Y_{1},\ldots,Y_{m})$
is itself an i.i.d. sequence, $m\ge2$. In this case, we show that
the multiplicative gap $e/(e-1)$ in \eqref{eq:intro_hzbd} can be
replaced by
\begin{equation}
1+4.71\sqrt{\frac{\log m}{m}}.\label{eq:intro_gap_sid}
\end{equation}
This is proved in Proposition \ref{prop:sid_iid} and Theorem \ref{thm:sid_gap}.
This means that as $m$ increases, $(Y_{1},\ldots,Y_{m})$ becomes
closer to being SID in a uniform manner (that does not depend on the
distribution of $Y_{i}$). This can be regarded as an information
analogue of Kolmogorov's uniform theorem. We conjecture that the optimal
multiplicative gap is $1+O(1/\sqrt{m})$, that is, the $\log m$ term
can be eliminated.

Theorem \ref{thm:sid_gap} can also be regarded as a one-sided variant
of Kolmogorov's uniform theorem, where we require the infinitely divisible
estimate to stochastically dominates the original distribution. This
may be of independent interest outside of information theory. The
main challenge in proving Theorem \ref{thm:sid_gap}, compared to
Kolmogorov's uniform theorem, is that the stochastic dominance condition
is significantly more stringent than having a small $d_{\mathrm{U}}$
(which is a rather weak condition on the tails of the distributions).

We list some potential applications of our result.\bigskip{}

\subsection{Independent Component Analysis and Blind Source Separation\label{subsec:iica_intro}}

Independent component analysis (ICA) \cite{jutten1991blind,comon1994independent,hyvarinen2000independent}
concerns the problem of recovering independent signals from their
observed mixture. While assumptions on the distributions of the individual
signals are not usually imposed, the model (under which signals are
mixed to form the observation) is usually assumed to take a certain
prescribed form (e.g. the signals are summed together linearly). 

The generalized ICA setting \cite{painsky2014generalized,painsky2015generalized}
(also see \cite{barlow1989finding,himberg2001independent,yeredor2007ica,yeredor2011independent,gutch2012ica,vsingliar2006noisy,nguyen2011binary}
for related problems) removes the assumption on the model. Given the
observation $X=(Y_{1},\ldots,Y_{n})$, $Y_{i}\in\{1,\ldots,a\}$,
the generalized ICA aims at finding the signals $Z_{1},\ldots,Z_{n}\in\{1,\ldots,a\}$
and a bijective mapping $f:\{1,\ldots,a\}^{n}\to\{1,\ldots,a\}^{n}$
such that $X=f(Z_{1},\ldots,Z_{n})$, and $Z_{1},\ldots,Z_{n}$ are
as mutually independent as possible, which is accomplished by minimizing
the total correlation \cite{watanabe1960information}
\[
C(Z_{1};\ldots;Z_{n}):=\sum_{i=1}^{n}H(Z_{i})-H(Z_{1},\ldots,Z_{n}).
\]
The generalized ICA does not make any assumption on the mapping $f$
other than that it is bijective. In \cite{painsky2017large}, it is
shown that if $a=2$ and the distribution of $X$ is generated uniformly
over the probability simplex, then the average optimal $C(Z_{1};\ldots;Z_{n})$
can be upper-bounded by a constant.  The main differences between
the generalized ICA and our result \eqref{eq:intro_hzbd} are that
we require $Z_{1},\ldots,Z_{n}$ to be exactly (not only approximately)
mutually independent and have the same distribution, but $f$ does
not need to be bijective, and we do not make any assumption on the
size of the alphabet of $Z_{i}$.

Our result can be regarded as a variant of the generalized ICA, which
we call independent identically-distributed component analysis (IIDCA).
Suppose $Z_{1},\ldots,Z_{n}$ are i.i.d. following an unknown distribution,
and $f(z_{1},\ldots,z_{n})$ is an unknown (not necessarily injective)
function. We are able to estimate the distribution of $X=f(Z_{1},\ldots,Z_{n})$
(e.g. by i.i.d. samples of $X$). The goal is to learn the distribution
of $Z_{i}$ and the function $f$ (since the labeling of the values
of $Z_{i}$ is lost, we can only learn the entries of the pmf of $Z_{i}$,
but not the actual values of $Z_{i}$). In ICA, the components are
mixed together via addition, which causes loss of information. In
IIDCA, the loss of information is due to the unknown nature of the
function (or the loss of the labeling), and also that $f$ may not
be injective.

One approach to IIDCA is to solve the following optimization problem:
minimize $H(Z_{1})$ subject to the constraint that $Z_{1},\ldots,Z_{n}$
are i.i.d., and $H(X|Z_{1},\ldots,Z_{n})=0$. This would minimize
the amount of information lost by the function, i.e., $H(Z_{1},\ldots,Z_{n})-H(X)$.
By \eqref{eq:intro_hzbd}, we can always achieve $H(Z_{1})\le(e/(e-1))H(X)/n+2.43$.
Refer to Section \ref{sec:iica} for an approximate algorithm.

For a concrete example, consider the compression of a $5\times5$
image, where the color of each pixel $Z_{i}$ ($i=1,\ldots,25$) is
assumed to be i.i.d. following an unknown distribution $p_{Z}$.
We observe the image through a deterministic but unknown transformation
$f(z_{1},\ldots,z_{25})$ (which is an arbitrary distortion of the
image that does not need to preserve locations and colors of pixels).
 We observe a number of i.i.d. samples of transformed images (the
transformation $f$ is the same for all samples). The goal is to estimate
the distribution of colors $p_{Z}$ and the transformation $f$. Since
$f$ is arbitrary, we can only hope for learning the entries of the
pmf $p_{Z}$, but not which color these entries correspond to. The
IIDCA provides a method to recover the distribution of $Z_{i}$ (up
to permutation) without any assumption on the transformation $f$.

\bigskip{}

\subsection{Distributed Storage with a Secrecy Constraint}

Suppose we would divide a piece of data $X$ (a discrete random variable)
into $n$ pieces $Z_{1},\ldots,Z_{n}$, where each piece is stored
in a separate node. There are three requirements:
\begin{enumerate}
\item (Recoverability) $X$ is recoverable from the pieces of all $n$ nodes,
i.e., $H(X|Z_{1},\ldots,Z_{n})=0$.
\item (Mutual secrecy) Each node has no information about the pieces stored
at other nodes, i.e., $Z_{1},\ldots,Z_{n}$ are mutually independent.
\item (Identity privacy) Each node has no information about which piece
of data it stores, i.e., if $J\sim\mathrm{Unif}\{1,\ldots,n\}$, then
we require that $J$ is independent of $Z_{J}$. This meaning that
if we assign the pieces to the nodes randomly, then a node which stores
$Z_{J}$ (let $J$ be the index of the piece it stores) cannot use
$Z_{J}$ to gain any information about $J$. This condition is equivalent
to that $Z_{1},\ldots,Z_{n}$ have the same distribution.
\end{enumerate}
Our goal is to minimize the amount of storage needed at each node,
i.e., $\max_{1\le i\le n}H(Z_{i})$. These three requirements are
trivially achievable if $X=(Y_{1},\ldots,Y_{m})$ contains $m$ i.i.d.
components, where $m$ is divisible by $n$, since we can simply let
$Z_{i}=(Y_{(i-1)m/n+1},\ldots,Y_{im/n})$. Nevertheless, this strategy
fails when the data does not follow such homogeneous structure. For
example, assume $X=(Y_{1},\ldots,Y_{100})$ contains the net worth
of $100$ individuals, and $Y_{1}$ corresponds to a well-known billionaire
(and thus has a much higher expectation than other $Y_{i}$'s). If
we simply divide the dataset into 10 pieces (each with 10 data points)
to be stored in 10 nodes, then identity privacy is not guaranteed,
and if a node observes a data point much larger than the rest, it
will be able to identify the billionaire and know the billionaire's
net worth, which may be considered a privacy breach. Hence, identity
privacy is critical to the privacy of the user's data.

The construction in \eqref{eq:intro_hzbd} provides a way to compute
$Z_{1},\ldots,Z_{n}$ satisfying the three requirements, and requires
an amount of storage $\max_{i}H(Z_{i})\le(e/(e-1))H(X)/n+2.43$ at
each node, a multiplicative factor $e/(e-1)\approx1.582$ larger than
$H(X)/n$. An advantage of our construction is that it does not require
$X$ to follow any structure. If $X=(Y_{1},\ldots,Y_{m})$ indeed
follows an i.i.d. structure, then we can apply \eqref{eq:intro_gap_sid}
to reduce the storage requirement, even when $m<n$ (note that the
aforementioned strategy $Z_{i}=(Y_{(i-1)m/n+1},\ldots,Y_{im/n})$
requires $m\ge n$).

We remark that even if  the identity privacy requirement is removed,
this is still a non-trivial problem, and \eqref{eq:intro_gap_sid}
still provides a reasonable construction.

This setting is related to the secret sharing scheme \cite{shamir1979share}
and wiretap channel II \cite{ozarow1984wire}, where the data is recoverable
from the pieces in any $k$ nodes, but is secret from an eavesdropper
that can access an arbitrary subset of $k'<k$ nodes. See \cite{chor1995private,shah2011information,goparaju2013data}
for other settings on distributed storage with privacy or secrecy
constraints. Note that these previous works assume that $X$ follows
a uniform distribution (over bit sequences or sequences of symbols
in a finite field), and does not concern the case where $X$ is an
arbitrary discrete random variable.\bigskip{}

\subsection{Distributed Random Number Generation}

Suppose we want to generate a discrete random variable $X$ following
the distribution $p_{X}$, using $N$ identical and independent random
number generators. Each generator is capable of generating a random
variable following the distribution $p_{Z}$. We are allowed to design
$p_{Z}$ according to $p_{X}$. Nevertheless, some of the generators
may fail, and we want to guarantee that $X$ can be generated whenever
$n$ out of the $N$ generators are working.

One strategy is to let $p_{Z}$ be $\mathrm{Unif}\{1,\ldots,2^{k}\}$
(i.e., $k$ i.i.d. fair bits), and use a discrete distribution generating
tree \cite{knuth1976complexity} to generate $X$ (traverse the tree
according to the bits produced by the first working generator, then
the second working generator, and so on). We declare failure if a
leaf node is not reached after we exhaust all working generators.
This strategy would introduce a bias to the distribution of $X$,
since the less probable values of $X$ (corresponding to leave nodes
of larger depth) are less likely to be reached. Therefore, this strategy
is undesirable if we want $X$ to follow $p_{X}$ exactly. Therefore,
a fixed length scheme (which declares failure if there are fewer than
$n$ working generators, where $n$ is a fixed threshold) is preferrable
over a variable length scheme.

Using the construction in \eqref{eq:intro_hzbd}, we can design $p_{Z}=p_{Z_{i}}$
such that $X$ can be generated by the output of any $n$ generators.
We declare failure if there are fewer than $n$ working generators.
We can guarantee that $X\sim p_{X}$ conditional on the event that
there is no failure. The entropy used per generator is bounded by
$(e/(e-1))H(X)/n+2.43$.

\bigskip{}

\subsection{Bridge Between One-shot and IID Results}

While most results in information theory (e.g. channel coding theorem,
lossy source coding theorem) are proved in the i.i.d. asymptotic regime,
one-shot results are also widely studied, with Huffman coding \cite{huffman1952method}
(a one-shot lossless source coding result) being a notable example.
See \cite{shannon1959coding,goblick1963coding,posner1971epsilon,kostina2012fixed,palzer2016converse,kostina2016nonasymptotic,sfrl_trans,elkayam2020one}
for results on one-shot or finite-blocklength lossy source coding,
and \cite{feinstein1954new,shannon1957certain,blackwell1959capacity,gallager1965simple,hayashi2003general,polyanskiy2010channel,verdu2012nonasymp,yassaee2013oneshot,li2019unified}
for results on one-shot channel coding.

The results in this paper allows us to discover near-i.i.d. structures
in one-shot settings, without explicitly assuming it. For example,
in a one-shot source coding setting where we would compress the random
variable $X$, we can apply \eqref{eq:intro_hzbd} to divide $X$
into i.i.d. $Z_{1},\ldots,Z_{n}$, and then apply standard techniques
on i.i.d. source such as the method of type \cite{shannon1948mathematical,cover1999elements}.
Therefore, the results in this paper can provide a bridge between
one-shot and i.i.d. results. While \eqref{eq:intro_hzbd} suffers
from a multiplicative penalty, the penalty may be reduced if $X$
follows certain structure (e.g. \eqref{eq:intro_gap_sid}). Similar
approaches have been used in \cite{barlow1989finding,painsky2017large}
for large alphabet source coding.

One particular structure that allows a simple i.i.d. representation
is Markov chain. Suppose we would compress $X=(Y_{1},\ldots,Y_{m})$,
where $Y_{1},\ldots,Y_{m}$ forms a Markov chain (see \cite{vavsek1980error,davisson1981error,natarajan1985large}
for results on source coding for Markov chains). We may represent
$Y_{i+1}=f(Y_{i},W_{i})$ as a function of $Y_{i}$ and $W_{i}$,
where $W_{i}$ is a random variable independent of $\{Y_{j}\}_{j\le i},\{W_{j}\}_{j<i}$.
This is called the functional representation lemma \cite{elgamal2011network}
or the innovation representation \cite{painsky2019innovation}. We
can assume $W_{1},\ldots,W_{m-1}$ to be i.i.d. if the Markov chain
is time-homogeneous. Therefore, to compress $Y_{1},\ldots,Y_{m}$,
it suffices to compress $Y_{1}$ and the i.i.d. sequence $W_{1},\ldots,W_{m-1}$.
The minimum of $H(W_{i})$ has been studied in \cite{vidyasagar2012metric,kovavcevic2015entropy,kocaoglu2017greedy,cicalese2019minimum,li2020efficient}.\bigskip{}

\subsection{Other Concepts of Divisibility in Information Theory}

We remark that continuous-time memoryless channels, such as the continuous-time
additive white Gaussian noise channel \cite{shannon1948mathematical}
and the Poisson channel \cite{kabanov1978capacity,davis1980capacity,wyner1988capacity},
can be regarded as ``infinitely divisible'' since one can break
the time interval into segments. Nevertheless, the existence of infinitely
divisible channels does not imply the existence of informationally
infinitely divisible discrete random variables. Information-theoretic
analysis on infinitely divisible distributions (Poisson distribution
in particular) has been performed in, e.g. \cite{kontoyiannis2004entropy,kontoyiannis2005entropy,yu2009entropy,harremoes2010thinning,barbour2010compound,johnson2013log},
though they concerned the classical definition of infinite divisibility,
not the ``informational infinite divisibility'' in this paper. A
marked Poisson process \cite{haenggi2012stochastic} can be considered
as an informationally infinitely divisible random object. More precisely,
let $\{(X_{i},T_{i})\}_{i}$ be a point process, where $\{T_{i}\}_{i}$
is a Poisson process over $[0,1]$, and $X_{i}\stackrel{iid}{\sim}p_{X}$.
The point process $\{(X_{i},T_{i})\}_{i}$ is informationally infinitely
divisible since we can divide it into $n$ i.i.d. pieces, where the
$k$-th piece is $\{(X_{i},T_{i}-(k-1)/n):\,(k-1)/n\le T_{i}<kn\}$.
However, a marked Poisson process is not discrete and has infinite
entropy (since the times $T_{i}$ are continuous). Nevertheless, our
construction in Theorem \ref{thm:pin_gap}, loosely speaking, can
be regarded as a marked Poisson process where the information about
the times $T_{i}$ are removed. Marked Poisson processes have also
been used in information theory to construct coding schemes in \cite{sfrl_trans,li2019unified}.
Successive refinement in source coding \cite{equitz1991successive}
is sometimes referred as source divisibility (e.g. \cite{koshelev1994divisibility,haroutunian2000successive}),
though this division is performed in a heterogeneous manner (each
description is intended for a different distortion level), not in
the homogeneous i.i.d. manner in this paper.

\bigskip{}

\subsection*{Notations}

Throughout this paper, we assume that the entropy $H$ is in bits,
$\log$ is to base $2$, and $\ln$ is to base $e$. The binary entropy
function is $H_{\mathrm{b}}(x):=-x\log x-(1-x)\log(1-x)$. We write
$\mathbb{N}=\{1,2,3,\ldots\}$, $[n]=\{1,\ldots,n\}$. The probability
mass function (pmf) of a random variable $X$ is denoted as $p_{X}$.
The support of a probability mass function $p$ is denoted as $\mathrm{supp}(p)$.
For a pmf $p$ over the set $\mathcal{X}$, and a pmf $q$ over the
set $\mathcal{Y}$, the product pmf $p\times q$ is a pmf over $\mathcal{X}\times\mathcal{Y}$
with $(p\times q)(x,y):=p(x)q(y)$. Write $p^{\times n}:=p\times\cdots\times p$
($n$ terms on the right hand side). For two random variables $X$
and $Y$, $X\stackrel{d}{=}Y$ means that $X$ and $Y$ have the same
distribution.

The cumulative distribution function (cdf) of a random variable $X$
is denoted as $F_{X}$. The space of cdf's over $[0,\infty)$ is denoted
as $\mathcal{P}_{+}$. For a cdf $F\in\mathcal{P}_{+}$, the inverse
cdf is given by $F^{-1}(t)=\inf\{x\ge0:\,F(x)\ge t\}$. For two cdf's
$F_{1},F_{2}$, we say $F_{1}$ stochastically dominates $F_{2}$,
written as $F_{1}\le F_{2}$, if $F_{1}(t)\le F_{2}(t)$ for any $t$.
The convolution of two cdf's $F_{1},F_{2}\in\mathcal{P}_{+}$ is given
by $(F_{1}*F_{2})(t):=\int_{0}^{t}F_{1}(t-s)dF_{2}(s)$. Write $F^{*n}:=F*\cdots*F$
($n$ terms on the right hand side). The uniform metric between cdf's
is denoted as $d_{\mathrm{U}}(F_{1},F_{2}):=\sup_{x}|F_{1}(x)-F_{2}(x)|$.
The total variation distance is denoted as $d_{\mathrm{TV}}(F_{1},F_{2}):=\sup_{A\subseteq\mathbb{R}\,\mathrm{measurable}}|\int\mathbf{1}\{t\in A\}dF_{1}(t)-\int\mathbf{1}\{t\in A\}dF_{2}(t)|$.
The mean of the distribution given by the cdf $F\in\mathcal{P}_{+}$
is denoted as
\[
E(F):=\mathbf{E}_{T\sim F}[T]=\int_{0}^{\infty}tdF(t).
\]

The pmf of the Bernoulli distribution is denoted as $\mathrm{Bern}(x;\gamma):=\mathbf{1}\{x=0\}(1-\gamma)+\mathbf{1}\{x=1\}\gamma$.
The pmf of the geometric distribution over $\mathbb{N}$ is denoted
as $\mathrm{Geom}(x;\gamma):=\gamma(1-\gamma)^{x-1}$. The binomial,
negative binomial and Poisson distribution are denoted as $\mathrm{Bin}(x;\,n,p)$,
$\mathrm{NegBin}(x;\,r,p)$ and $\mathrm{Poi}(x;\lambda)$ respectively.
We use the currying notation $\mathrm{Bern}(\gamma)$ to denote the
pmf $x\mapsto\mathrm{Bern}(x;\gamma)$ (i.e., $\mathrm{Bern}(\gamma)$
is a function with $\mathrm{Bern}(\gamma)(x)=\mathrm{Bern}(x;\gamma)$).
Similar for $\mathrm{Geom}$ and other distributions.

\medskip{}
\medskip{}

\section{Information Spectrum\label{sec:spectrum}}

For a discrete random variable $X$ with pmf $p_{X}$, its \emph{self
information} is defined as $\iota_{X}(x):=-\log p_{X}(x)$. Its \emph{information
spectrum cdf} is defined as
\begin{align*}
F_{\iota_{X}}(t) & :=\mathbf{P}\left(\iota_{X}(X)\le t\right)\\
 & =\mathbf{P}\left(-\log p_{X}(x)\le t\right).
\end{align*}
Note that $H(X)=E(F_{\iota_{X}})=\int_{0}^{\infty}tdF_{\iota_{X}}(t)$
is the mean of $\iota_{X}(X)$, and if $X$ is independent of $Y$,
then the joint self information is $\iota_{(X,Y)}(x,y)=\iota_{X}(x)+\iota_{Y}(y)$,
and hence $F_{\iota_{(X,Y)}}(t)=(F_{\iota_{X}}*F_{\iota_{Y}})(t)=\int_{0}^{t}F_{\iota_{X}}(t-s)dF_{\iota_{Y}}(s)$.

Not every cdf over $[0,\infty)$ is the information spectrum cdf of
some discrete random variable. For a cdf over $[0,\infty)$ to be
the information spectrum cdf, that cdf must correspond to a discrete
distribution, where the probability of $t$ is a multiple of $2^{-t}$.
We define the set of information spectrum cdf's as
\[
\mathcal{P}_{\iota}:=\{F_{\iota_{X}}:\,X\text{ is discrete RV}\}.
\]

We present the concept of aggregation in \cite{vidyasagar2012metric,cicalese2016approximating}.
\begin{defn}
For two pmf's $p_{X},p_{Y}$, we say $p_{Y}$ is an \emph{aggregation}
of $p_{X}$, written as $p_{X}\sqsubseteq p_{Y}$, if there exists
a function $g:\mathrm{supp}(p_{X})\to\mathrm{supp}(p_{Y})$ (called
the \emph{aggregation map}) such that $p_{Y}$ is the pmf of $g(X)$,
where $X\sim p_{X}$.
\end{defn}
Also recall the concept of majorization over pmf's (see \cite{marshall1979inequalities}):
\begin{defn}
For two pmf's $p_{X},p_{Y}$, we say $p_{X}$ is \emph{majorized by}
$p_{Y}$, written as $p_{X}\preceq p_{Y}$, if 
\begin{equation}
\max_{A\subseteq\mathrm{supp}(p):\,|A|\le k}p_{X}(A)\le\max_{B\subseteq\mathrm{supp}(q):\,|B|\le k}p_{Y}(B)\label{eq:maj}
\end{equation}
for any $k\in\mathbb{N}$ (write $p_{X}(A):=\sum_{x\in A}p_{X}(x)$).
Equivalently, the sum of the $k$ largest $p_{X}(x)$'s is not greater
than the sum of the $k$ largest $p_{Y}(x)$'s.
\end{defn}
It is shown in \cite{cicalese2016approximating} that $p_{X}\sqsubseteq p_{Y}$
implies $p_{X}\preceq p_{Y}$. In fact, it is straightforward to check
that (refer to \eqref{eq:F_inv_logGp} and Proposition \ref{prop:imaj_maj}
for a proof)
\[
p_{X}\sqsubseteq p_{Y}\,\Rightarrow\,F_{\iota_{X}}\le F_{\iota_{Y}}\,\Rightarrow\,p_{X}\preceq p_{Y}.
\]
For the other direction, it is shown in \cite{li2020efficient} that
$p_{X}\preceq p_{Y}$ implies 
\begin{equation}
p_{X}\times\mathrm{Geom}(1/2)\sqsubseteq p_{Y},\label{eq:geom_ag}
\end{equation}
where $p_{X}\times\mathrm{Geom}(1/2)$ is the joint pmf of $(X,Z)$
where $X\sim p_{X}$ is independent of $Z\sim\mathrm{Geom}(1/2)$.

We now generalize the concept of majorization to arbitrary cdf's over
$[0,\infty)$.
\begin{defn}
For two cdf's $F_{1},F_{2}\in\mathcal{P}_{+}$, we say that $F_{1}$
is \emph{informationally majorized} by $F_{2}$, written as $F_{1}\stackrel{\iota}{\preceq}F_{2}$,
if $G_{F_{1}}(\gamma)\ge G_{F_{2}}(\gamma)$ for any $\gamma\in[0,1]$,
where $G_{F}:[0,1]\to[0,\infty)$, 
\[
G_{F}(\gamma):=\int_{-\infty}^{\infty}2^{t}d\min\{F(t),\,\gamma\},
\]
where $F(t)=0$ for $t<0$.
\end{defn}
It is clear that ``$\stackrel{\iota}{\preceq}$'' is a transitive
relation. It is straightforward to check that $G_{F}(\gamma)$ is
continuous, strictly increasing, convex, and $G_{F}(\gamma)\ge\gamma$.
Write $G_{F}'$ for the left derivative of $G_{F}$ (let $G_{F}'(0)=0$).
Since
\begin{align*}
G_{F}(\gamma) & =\int_{-\infty}^{\infty}2^{t}d\min\{F(t),\,\gamma\}\\
 & =\int_{0}^{\gamma}2^{F^{-1}(s)}ds,
\end{align*}
we have
\begin{equation}
F^{-1}(\gamma)=\log G_{F}'(\gamma).\label{eq:F_inv_logGp}
\end{equation}
As a result, $F_{1}\le F_{2}$  implies $F_{1}\stackrel{\iota}{\preceq}F_{2}$.

We first show that informational majorization is a generalization
of majorization. The proof is given in Appendix \ref{subsec:pf_imaj_maj}.
\begin{prop}
\label{prop:imaj_maj}For two discrete random variables $X,Y$, we
have $p_{X}\preceq p_{Y}$ if and only if $F_{\iota_{X}}\stackrel{\iota}{\preceq}F_{\iota_{Y}}$.
\end{prop}
\medskip{}
The mean $E(F)=\int_{0}^{\infty}tdF(t)$ is non-increasing with respect
to ``$\stackrel{\iota}{\preceq}$'', as shown in the following proposition.
This is an analogue of the fact that entropy is Schur-concave. The
proof is given in Appendix \ref{subsec:pf_e_noninc}.
\begin{prop}
\label{prop:e_noninc}If $F_{1}\stackrel{\iota}{\preceq}F_{2}$, then
$E(F_{1})\ge E(F_{2})$.
\end{prop}
\medskip{}
Convolution and mixture preserves ``$\stackrel{\iota}{\preceq}$'',
as shown in the following proposition. The proof is given in Appendix
\ref{subsec:pf_conv_imaj}.
\begin{prop}
\label{prop:conv_imaj}If $F_{1}\stackrel{\iota}{\preceq}F_{2}$,
$F_{3}\stackrel{\iota}{\preceq}F_{4}$, then $F_{1}*F_{3}\stackrel{\iota}{\preceq}F_{2}*F_{4}$,
and $(1-\lambda)F_{1}+\lambda F_{3}\stackrel{\iota}{\preceq}(1-\lambda)F_{2}+\lambda F_{4}$
for any $\lambda\in[0,1]$.
\end{prop}
\medskip{}

Any cdf is closely approximated by an information spectrum cdf, as
shown in the following proposition.
\begin{prop}
\label{prop:infcdf_approx}For any cdf $F\in\mathcal{P}_{+}$, there
exists a discrete random variable $X$ such that $F_{\iota_{X}}\stackrel{\iota}{\preceq}F$
and
\begin{align*}
E(F)\le H(X) & \le E(F)+H_{\mathrm{b}}\left(\min\left\{ \sqrt{\frac{E(F)}{\log e}},\,\frac{1}{2}\right\} \right)\\
 & \le E(F)+1.
\end{align*}
\end{prop}
\begin{IEEEproof}
Let $G=G_{F}$. Let $G'(\gamma)$ be the left derivative of $G$.
Let $G^{-1}(t)$ be the inverse function of $G$ (let $G^{-1}(t)=1$
if $t\ge\sup_{\gamma}G(\gamma)$), and let $\gamma_{k}:=G^{-1}(k)$
for $k\ge0$, and $\gamma_{-1}:=-1$. By the convexity of $G$, $\gamma_{k}$
is a concave sequence. Let $q_{k}:=\gamma_{k}-\gamma_{k-1}$ for $k\ge0$
(note that $q_{0}=1$). By the concavity of $\gamma_{k}$, $q_{k}$
is a non-increasing sequence.

Let $X$ be a random variable with $p_{X}(k)=q_{k}$ for $k\ge1$.
Write $\tilde{G}=G_{F_{\iota_{X}}}$. Note that $\tilde{G}(\gamma_{k})=G(\gamma_{k})=k$
for any $k\ge0$, and that $\tilde{G}(\gamma)$ is affine over the
interval $[\gamma_{k-1},\,\gamma_{k}]$. By the convexity of $G$,
we have $\tilde{G}(\gamma)\ge G(\gamma)$ for any $\gamma$, and hence
$F_{\iota_{X}}\stackrel{\iota}{\preceq}F$. The lower bound in the
proposition is a consequence of Proposition \ref{prop:e_noninc}.
To prove the upper bound,
\begin{align*}
E(F) & =\int_{0}^{\infty}tdF(t)\\
 & \stackrel{(a)}{=}\int_{0}^{1}\log G'(\gamma)d\gamma\\
 & \stackrel{(b)}{\ge}\int_{q_{1}}^{1}\log\frac{1}{q_{1}}d\gamma\\
 & =(1-q_{1})\log\frac{1}{q_{1}}\\
 & \ge(1-q_{1})^{2}\log e,
\end{align*}
where (a) is due to \eqref{eq:F_inv_logGp}, and (b) is is because
$G'(\gamma)$ is non-decreasing, and $1/q_{1}$ is the slope of the
chord of $G$ between $0$ and $q_{1}$. Hence,
\[
q_{1}\ge1-\sqrt{\frac{E(F)}{\log e}}.
\]
We have
\begin{align*}
E(F) & =\int_{0}^{\infty}tdF(t)\\
 & =\int_{0}^{1}F^{-1}(\gamma)d\gamma\\
 & =\int_{0}^{1}\log G'(\gamma)d\gamma\\
 & =\sum_{k=1}^{\infty}\int_{\gamma_{k-1}}^{\gamma_{k}}\log G'(\gamma)d\gamma\\
 & \stackrel{(a)}{\ge}\sum_{k=1}^{\infty}(\gamma_{k}-\gamma_{k-1})\log\frac{1}{\gamma_{k-1}-\gamma_{k-2}}\\
 & =\sum_{k=1}^{\infty}q_{k}\log\frac{1}{q_{k-1}}\\
 & =H(X)+\sum_{k=1}^{\infty}q_{k}\log\frac{q_{k}}{q_{k-1}}\\
 & \stackrel{(b)}{\ge}H(X)+q_{1}\log q_{1}+\left(\sum_{k=2}^{\infty}q_{k}\right)\log\frac{\left(\sum_{k=2}^{\infty}q_{k}\right)}{\left(\sum_{k=2}^{\infty}q_{k-1}\right)}\\
 & =H(X)+q_{1}\log q_{1}+(1-q_{1})\log(1-q_{1})\\
 & \ge H(X)-H_{\mathrm{b}}\left(\min\left\{ \sqrt{\frac{E(F)}{\log e}},\,\frac{1}{2}\right\} \right),
\end{align*}
where (a) is because $G'(\gamma)$ is non-decreasing, and $1/(\gamma_{k-1}-\gamma_{k-2})$
is the slope of the chord of $G$ between $\gamma_{k-2}$ and $\gamma_{k-1}$
(for $k=1$, $1/(\gamma_{k-1}-\gamma_{k-2})=1\le G'(\gamma)$), and
(b) is due to the log sum inequality.

\end{IEEEproof}
\medskip{}

\section{Multiplicative Gap to Infinite Divisibility}

Define the space of $n$-divisible cdf's $\mathcal{P}_{+}^{*n}$ and
the space of infinitely divisible cdf's $\mathcal{P}_{+}^{*\infty}$
as
\begin{align*}
\mathcal{P}_{+}^{*n} & :=\left\{ F^{*n}:\,F\in\mathcal{P}_{+}\right\} ,\\
\mathcal{P}_{+}^{*\infty} & :=\bigcap_{n\in\mathbb{N}}\mathcal{P}_{+}^{*n}.
\end{align*}
Similarly, define the space of $n$-informationally-divisible cdf's
$\mathcal{P}_{\iota}^{*n}$ and the space of informationally infinitely
divisible cdf's $\mathcal{P}_{\iota}^{*\infty}$ as
\begin{align*}
\mathcal{P}_{\iota}^{*n} & :=\left\{ F^{*n}:\,F\in\mathcal{P}_{\iota}\right\} \\
\mathcal{P}_{\iota}^{*\infty} & :=\bigcap_{n\in\mathbb{N}}\mathcal{P}_{\iota}^{*n}.
\end{align*}
Note that $\mathcal{P}_{\iota}^{*n}$ is the set of $F_{\iota_{Z^{n}}}$
for i.i.d. sequences $Z^{n}=(Z_{1},\ldots,Z_{n})$. A random variable
$X$ is informationally infinitely divisible (as described in the
introduction) if and only if $F_{\iota_{X}}\in\mathcal{P}_{\iota}^{*\infty}$.

The following proposition shows that there is no informationally infinitely
divisible random variable.
\begin{prop}
$\mathcal{P}_{\iota}^{*\infty}=\emptyset$.
\end{prop}
\begin{IEEEproof}
Consider i.i.d. discrete random variables $Z^{n}=(Z_{1},\ldots,Z_{n})$,
$Z_{i}\sim p_{Z}$, where we assume $Z\in\mathbb{N}$, $p_{Z}(1)\ge p_{Z}(2)\ge\cdots$.
Let $a$ be the largest integer such that $p_{Z}(a)=p_{Z}(1)$. Assume
$Z$ is not a uniform random variable (i.e., $p_{Z}(a+1)>0$). Consider
the pmf $p_{Z^{n}}$. Its largest entries have values $(p_{Z}(1))^{n}$,
and its second largest entries have values $(p_{Z}(1))^{n-1}p_{Z}(a+1)$.
The number of entries $z^{n}\in\mathbb{N}^{n}$ with value $p_{Z^{n}}(z^{n})=(p_{Z}(1))^{n-1}p_{Z}(a+1)$
is a multiple of $n$ (since it is the number of vectors $z^{n}\in\mathbb{N}^{n}$
such that exactly one $z_{i}$ has $p_{Z}(z_{i})=p_{Z}(a+1)$, and
every other $z_{i}$'s has $p_{Z}(z_{i})=p_{Z}(1)$, so the number
of such vectors is a multiple of $n$ by considering which component
has $p_{Z}(z_{i})=p_{Z}(a+1)$). Therefore, either $Z$ is uniform,
or the number of second largest entries of $p_{Z^{n}}$ is a multiple
of $n$.

Assume the contrary that there exists random variable $X$ such that
$\mathcal{F}_{\iota_{X}}\in\mathcal{P}_{\iota}^{*\infty}$. It is
clear that $X$ cannot be uniform (by taking $n$ to be larger than
its cardinality). Consider the number of second largest entries of
$p_{X}$, and let $n$ to be larger than this number. Since $\mathcal{F}_{\iota_{X}}\in\mathcal{P}_{\iota}^{*n}$,
there exists i.i.d. discrete random variables $Z^{n}=(Z_{1},\ldots,Z_{n})$
such that $\mathcal{F}_{\iota_{X}}=\mathcal{F}_{\iota_{Z^{n}}}$.
Since $X$ is not uniform, $Z_{i}$ cannot be uniform, and hence the
number of second largest entries of $p_{Z^{n}}$ is a multiple of
$n$, which gives a contradiction.
\end{IEEEproof}
\medskip{}

Nevertheless, there is a bounded multiplicative gap between $\mathcal{P}_{\iota}$
and $\mathcal{P}_{+}^{*\infty}$. In fact, there is a bounded multiplicative
gap between any cdf and $\mathcal{P}_{+}^{*\infty}$, as shown in
the following theorem.
\begin{thm}
\label{thm:infdiv_gap}For any cdf $F\in\mathcal{P}_{+}$ and $n\in\mathbb{N}\cup\{\infty\}$,
there exists $\tilde{F}\in\mathcal{P}_{+}^{*n}$ such that $\tilde{F}\le F$
and 
\[
E(F)\le E(\tilde{F})\le\frac{1}{1-(1-n^{-1})^{n}}E(F),
\]
where we assume $1/(1-(1-n^{-1})^{n})=e/(e-1)$ when $n=\infty$.
\end{thm}
\begin{IEEEproof}
Fix $\zeta\in(0,1)$. Let $U\sim\mathrm{Unif}[\zeta,1]$, and $Z:=F^{-1}(U)-F^{-1}(\zeta)$.
Let $Z_{1},Z_{2},\ldots$ be i.i.d. copies of $Z$. Let $N\sim\mathrm{Bin}(n,1-\zeta^{1/n})$
if $n<\infty$, and $N\sim\mathrm{Poi}(-\ln\zeta)$ if $n=\infty$,
independent of $\{Z_{i}\}$, and $V:=F^{-1}(\zeta)+\sum_{i=1}^{N}Z_{i}$.
Note that when $n=\infty$, $V$ is an infinitely divisible random
variable since it follows a compound Poisson distribution (with a
constant offset). We have $F_{V}\in\mathcal{P}_{+}^{*n}$. Let $V_{2}:=F^{-1}(\zeta)+\mathbf{1}\{N\ge1\}Z_{1}$
(note that $\mathbf{P}(N\ge1)=1-\zeta$), $V_{3}:=\mathbf{1}\{N=0\}F^{-1}(\tilde{U})+\mathbf{1}\{N\ge1\}(Z_{1}+F^{-1}(\zeta))$,
where $\tilde{U}\sim\mathrm{Unif}[0,\zeta]$. It is clear that $V_{3}\sim F$
and $V\ge V_{2}\ge V_{3}$, and hence $F_{V}\le F_{V_{2}}\le F$.

We then bound $\mathbf{E}[V]=E(F_{V})$. We have
\begin{align*}
\mathbf{E}[V] & =F^{-1}(\zeta)+\mathbf{E}[N]\mathbf{E}[Z]\\
 & =F^{-1}(\zeta)+\frac{\mathbf{E}[N]}{1-\zeta}\int_{\zeta}^{1}\left(F^{-1}(u)-F^{-1}(\zeta)\right)du\\
 & =\left(1-\mathbf{E}[N]\right)F^{-1}(\zeta)+\frac{\mathbf{E}[N]}{1-\zeta}\int_{\zeta}^{1}F^{-1}(u)du\\
 & \le\left(1-\mathbf{E}[N]\right)F^{-1}(\zeta)+\frac{\mathbf{E}[N]}{1-\zeta}E(F).
\end{align*}
The result follows from substituting $\zeta=(1-n^{-1})^{n}$ if $n<\infty$,
and $\zeta=e^{-1}$ if $n=\infty$, which makes $\mathbf{E}[N]=1$.
\end{IEEEproof}
\medskip{}

As a result, we can approximately divide any discrete random variable
into equal parts. The following theorem bounds the gap between $\mathcal{P}_{\iota}$
and $\mathcal{P}_{\iota}^{*n}$.
\begin{thm}
\label{thm:pin_gap}For any discrete random variable $X$ and $n\in\mathbb{N}$,
there exists i.i.d. random variables $Z_{1},\ldots,Z_{n}$ and function
$f$ such that $X\stackrel{d}{=}f(Z_{1},\ldots,Z_{n})$, and
\begin{align*}
 & H(Z_{1})-\frac{1}{1-(1-n^{-1})^{n}}\cdot\frac{H(X)}{n}\\
 & \le\min\left\{ 2.43,\,H_{\mathrm{b}}\left(\min\left\{ \sqrt{\frac{e}{(e-1)\log e}\cdot\frac{H(X)}{n}},\,\frac{1}{2}\right\} \right)+H_{\mathrm{b}}(2^{-1/n})+2(1-2^{-1/n})\right\} .
\end{align*}
\end{thm}
As a result, for any fixed $X$, we can achieve
\[
H(Z_{1})=\frac{e}{e-1}\cdot\frac{H(X)}{n}+O(n^{-1/2}\log n)
\]
as $n\to\infty$ (where the constant in $O(n^{-1/2}\log n)$ depends
on $H(X)$). We now prove the theorem.
\begin{IEEEproof}
We assume $n\ge2$ (the theorem is trivial for $n=1$). Fix any discrete
random variable $X$. By Theorem \ref{thm:infdiv_gap}, there exists
$\tilde{F}\in\mathcal{P}_{+}^{*n}$ such that $\tilde{F}\le F_{\iota_{X}}$
(and hence $\tilde{F}\stackrel{\iota}{\preceq}F_{\iota_{X}}$) and
\[
E(\tilde{F})\le\frac{1}{1-(1-n^{-1})^{n}}H(X).
\]
Let $F\in\mathcal{P}_{+}$ such that $F^{*n}=\tilde{F}$. By Proposition
\ref{prop:infcdf_approx}, there exists a discrete random variable
$Y$ such that $F_{\iota_{Y}}\stackrel{\iota}{\preceq}F$ and $H(Y)\le\int_{0}^{\infty}tdF(t)+H_{\mathrm{b}}(\min\{\sqrt{E(F)/\log e},\,1/2\})$.
By Proposition \ref{prop:conv_imaj}, $F_{\iota_{Y}}^{*n}\stackrel{\iota}{\preceq}F^{*n}=\tilde{F}\stackrel{\iota}{\preceq}F_{\iota_{X}}$.
Since $F_{\iota_{Y}}^{*n},F_{\iota_{X}}\in\mathcal{P}_{\iota}$, by
Proposition \ref{prop:imaj_maj}, $p_{Y}^{\times n}\preceq p_{X}$.
By \eqref{eq:geom_ag} (proved in \cite{li2020efficient}), $p_{Y}^{\times n}\times\mathrm{Geom}(1/2)\sqsubseteq p_{X}$.

Define a random variable $B\in\mathbb{N}$ as 
\begin{equation}
p_{B}(k):=(1-2^{-k})^{1/n}-(1-2^{-(k-1)})^{1/n}.\label{eq:pbdef}
\end{equation}
If $B_{1},\ldots,B_{n}$ are i.i.d. copies of $B$, then $\max\{B_{1},\ldots,B_{n}\}\sim\mathrm{Geom}(1/2)$.
Hence $p_{B}^{\times n}\sqsubseteq\mathrm{Geom}(1/2)$. Let $p_{Z}=p_{Y}\times p_{B}$.
We have $p_{Z}^{\times n}\sqsubseteq p_{Y}^{\times n}\times p_{B}^{\times n}\sqsubseteq p_{Y}^{\times n}\times\mathrm{Geom}(1/2)\sqsubseteq p_{X}$.
It is left to bound $H(Z)$. We have
\begin{align*}
H(Z) & =H(Y)+H(B)\\
 & \le E(F)+H_{\mathrm{b}}\left(\min\left\{ \sqrt{\frac{E(F)}{\log e}},\,\frac{1}{2}\right\} \right)+H(B)\\
 & =\frac{1}{n}E(\tilde{F})+H_{\mathrm{b}}\left(\min\left\{ \sqrt{\frac{E(\tilde{F})}{n\log e}},\,\frac{1}{2}\right\} \right)+H(B)\\
 & \le\frac{1}{1-(1-n^{-1})^{n}}\cdot\frac{H(X)}{n}+H_{\mathrm{b}}\left(\min\left\{ \sqrt{\frac{e}{(e-1)\log e}\cdot\frac{H(X)}{n}},\,\frac{1}{2}\right\} \right)+H(B).
\end{align*}
Note that
\begin{align*}
 & H_{\mathrm{b}}\left(\min\left\{ \sqrt{\frac{e}{(e-1)\log e}\cdot\frac{H(X)}{n}},\,\frac{1}{2}\right\} \right)\\
 & =O(n^{-1/2}\log n)
\end{align*}
as $n\to\infty$ for any fixed $H(X)$. It can be checked numerically
that $H(B)<1.43$ for $n\ge2$. For another bound on $H(B)$, since
the conditional distribution of $B$ given $B\ge2$ majorizes $\mathrm{Geom}(1/2)$,
we have
\begin{align*}
H(B) & \le H_{\mathrm{b}}(p_{B}(1))+(1-p_{B}(1))H(B|B\ge2)\\
 & \le H_{\mathrm{b}}(2^{-1/n})+2(1-2^{-1/n}).
\end{align*}
Note that $H_{\mathrm{b}}(2^{-1/n})+2(1-2^{-1/n})=O(n^{-1}\log n)$.
\end{IEEEproof}
\medskip{}

We remark that the condition ``$X\stackrel{d}{=}f(Z_{1},\ldots,Z_{n})$''
in Theorem \ref{thm:pin_gap} is equivalent to ``$H(X|Z_{1},\ldots,Z_{n})=0$''
in the introduction, since we can couple $Z_{1},\ldots,Z_{n}$ together
with $X$ such that $X=f(Z_{1},\ldots,Z_{n})$.\medskip{}

\section{Spectral Infinite Divisibility}

Although $\mathcal{P}_{\iota}^{*\infty}=\emptyset$ (i.e., there is
no informationally infinitely divisible random variable), we have
$\mathcal{P}_{\iota}\cap\mathcal{P}_{+}^{*\infty}\neq\emptyset$.
A discrete random variable $X$ is called \emph{spectral infinitely
divisible} (SID) if $F_{\iota_{X}}\in\mathcal{P}_{+}^{*\infty}$.
The distribution of $X$ is called a spectral infinitely divisible
distribution. A discrete uniform distribution is SID. Since a geometric
distribution is infinitely divisible, it is straightforward to check
that it is also SID. We now define a more general class of SID distributions
that includes uniform distributions and geometric distributions as
special cases.
\begin{defn}
Define the \emph{spectral negative binomial random variable} $X\sim\mathrm{SNB}(r,p,a,b)$
(where $r,a,b\in\mathbb{N}$, $p\in(0,1]$) over $\mathbb{N}$ as
follows: let
\[
s_{k}:=\binom{k+r-1}{r-1}ab^{k}
\]
for $k\in\mathbb{Z}_{\ge0}$, and
\[
p_{X}(x)=\mathrm{SNB}(x;\,r,p,a,b):=\frac{p^{r}}{a}\left(\frac{1-p}{b}\right)^{k},
\]
for $x\in\mathbb{N}$, where $k\in\mathbb{Z}_{\ge0}$ satisfies $\sum_{i=0}^{k-1}s_{k}<x\le\sum_{i=0}^{k}s_{k}$.
\end{defn}
It is straightforward to check that 
\[
\iota_{X}(X)\stackrel{d}{=}\log\frac{p^{r}}{a}+K\log\frac{1-p}{b},
\]
where $K$ follows the negative binomial distribution $\mathrm{NegBin}(r,p)$
with failure probability $p$ and number of failures until stopping
$r$. Since a negative binomial distribution is infinitely divisible,
the spectral negative binomial random variable is SID. The discrete
uniform distribution with cardinality $a$ is $\mathrm{SNB}(1,1,a,1)$.
The geometric distribution is $\mathrm{SNB}(1,p,1,1)$.

Another property is that if $X_{1}\sim\mathrm{SNB}(r_{1},p,a_{1},b)$
and $X_{2}\sim\mathrm{SNB}(r_{2},p,a_{2},b)$ are independent, then
there exists an injective function $f$ such that
\[
f(X_{1},X_{2})\sim\mathrm{SNB}(r_{1}+r_{2},\,p,\,a_{1}a_{2},\,b).
\]
\medskip{}

\section{Ratio to Infinite Divisibility}

More generally, we can measure how close a distribution is to infinite
divisibility as follows.
\begin{defn}
For a cdf $F\in\mathcal{P}_{+}$, define its \emph{ratio to infinite
divisibility} as
\[
r_{\mathrm{ID}}(F):=\frac{1}{E(F)}\inf\left\{ E(\tilde{F}):\,\tilde{F}\in\mathcal{P}_{+}^{*\infty},\,\tilde{F}\le F\right\} .
\]
If the mean $E(F)=0$, then $r_{\mathrm{ID}}(F):=1$.
\end{defn}
We list some properties of $r_{\mathrm{ID}}(F)$.
\begin{prop}
The ratio to infinite divisibility satisfies:
\begin{itemize}
\item (Bound) For any $F\in\mathcal{P}_{+}$,
\[
1\le r_{\mathrm{ID}}(F)\le\frac{e}{e-1}.
\]
\item (Relation to SID) A random variable $X$ is SID if and only if $r_{\mathrm{ID}}(F_{\iota_{X}})=1$.
\item (Convolution) For any $F_{1},F_{2}\in\mathcal{P}_{+}$ with positive
mean,
\[
r_{\mathrm{ID}}(F_{1}*F_{2})\le\frac{r_{\mathrm{ID}}(F_{1})E(F_{1})+r_{\mathrm{ID}}(F_{2})E(F_{2})}{E(F_{1})+E(F_{2})}.
\]
\end{itemize}
\end{prop}
\begin{IEEEproof}
The bound is a direct consequence of Theorem \ref{thm:infdiv_gap}.
For the relation to SID, the ``only if'' part is trivial. For the
``if'' part, if $r_{\mathrm{ID}}(F_{\iota_{X}})=1$, then there
exists $\tilde{F}_{i}\in\mathcal{P}_{+}^{*\infty}$, $i\in\mathbb{N}$
such that $\tilde{F}_{i}\le F_{\iota_{X}}$ and $\lim_{i\to\infty}E(\tilde{F}_{i})=E(F_{\iota_{X}})$.
Hence, $W_{1}(F_{\iota_{X}},\tilde{F}_{i})\to0$, where $W_{1}(F_{1},F_{2}):=\int_{-\infty}^{\infty}|F_{1}(t)-F_{2}(t)|dt$
is the 1-Wasserstein metric. Convergence in 1-Wasserstein metric implies
convergence in L\'evy metric, which is equivalent to weak convergence.
Since the limit of infinitely divisible distributions is infinitely
divisible, $F_{\iota_{X}}$ is infinitely divisible. The convolution
property follows from the fact that if $\tilde{F}_{i}\in\mathcal{P}_{+}^{*\infty}$,
$\tilde{F}_{i}\le F_{i}$, $i=1,2$, then $\tilde{F}_{1}*\tilde{F}_{2}\in\mathcal{P}_{+}^{*\infty}$
and $\tilde{F}_{1}*\tilde{F}_{2}\le F_{1}*F_{2}$.
\end{IEEEproof}
We can refine Theorem \ref{thm:pin_gap} using $r_{\mathrm{ID}}$
as follows. The proof is similar to Theorem \ref{thm:pin_gap} and
is omitted.
\begin{prop}
\label{prop:sid_iid}Fix any random variable $X$ and $n\in\mathbb{N}$.
There exists i.i.d. random variables $Z_{1},\ldots,Z_{n}$ and function
$f$ such that $X\stackrel{d}{=}f(Z_{1},\ldots,Z_{n})$, and
\begin{align*}
 & H(Z_{1})-r_{\mathrm{ID}}(F_{\iota_{X}})\frac{H(X)}{n}\\
 & \le\min\left\{ 2.43,\,H_{\mathrm{b}}\left(\min\left\{ \sqrt{\frac{e}{(e-1)\log e}\cdot\frac{H(X)}{n}},\,\frac{1}{2}\right\} \right)+H_{\mathrm{b}}(2^{-1/n})+2(1-2^{-1/n})\right\} .
\end{align*}
\end{prop}
\medskip{}

\section{Approximate Spectral Infinite Divisibility of IID Sequences}

In this section, we consider the case $X=Y^{m}=(Y_{1},\ldots,Y_{m})$,
where $Y_{1},\ldots,Y_{m}$ are i.i.d. random variables. As an information
analogue of Kolmogorov's uniform theorem \cite{kolmogorov1956two},
we show that $Y^{m}$ tends to being spectral infinitely divisible
uniformly, in the sense that $r_{\mathrm{ID}}(F_{\iota_{Y^{m}}})\to1$
as $m\to\infty$ uniformly. This result can be used together with
Proposition \ref{prop:sid_iid} to show that $Y^{m}$ can be divided
into $n$ i.i.d. random variables (with $n$ possibly larger than
$m$), with entropy close to the lower bound.

This result can also be regarded as a one-sided variant of Kolmogorov's
uniform theorem, where we require the infinitely divisible estimate
to stochastically dominates the original distribution, which may be
of independent interest.
\begin{thm}
\label{thm:sid_gap}For any cdf $F\in\mathcal{P}_{+}^{*m}$, $m\ge2$,
there exists $\tilde{F}\in\mathcal{P}_{+}^{*\infty}$ such that $\tilde{F}\le F$
and 
\[
E(\tilde{F})\le\left(1+4.71\sqrt{\frac{\log m}{m}}\right)E(F).
\]
As a result, for $m\ge2$,
\begin{equation}
\sup_{F\in\mathcal{P}_{+}^{*m}}r_{\mathrm{ID}}(F)\le1+4.71\sqrt{\frac{\log m}{m}}.\label{eq:th_rsid_bd}
\end{equation}

\medskip{}
\end{thm}
Theorem \ref{thm:sid_gap} can be stated in the following equivalent
way: For any i.i.d. random variables $T_{1},\ldots,T_{m}\ge0$, $m\ge2$,
there exists an infinitely divisible random variable $S$ such that
$S\ge\sum_{i=1}^{m}T_{i}$ almost surely, and
\[
\mathbf{E}\left[S-\sum_{i=1}^{m}T_{i}\right]\le4.71\sqrt{m\log m}\cdot\mathbf{E}[T_{1}].
\]

Before we prove the Theorem \ref{thm:sid_gap}, we first prove the
following lemma.
\begin{lem}
\label{lem:sid_gap_rec}If $\gamma\ge1$ and $m_{0}\in\mathbb{N}$
satisfy that 
\begin{equation}
\sup_{F\in\bigcup_{m\ge m_{0}}\mathcal{P}_{+}^{*m}}r_{\mathrm{ID}}(F)\le\gamma,\label{eq:lem_hyp}
\end{equation}
then for any $m>\max\{\gamma^{2}(m_{0}-2),\,0\}$,
\begin{equation}
\sup_{F\in\mathcal{P}_{+}^{*m}}r_{\mathrm{ID}}(F)\le\left(1+\frac{2\gamma}{m}\right)\left(2-\frac{1}{\gamma}\right).\label{eq:lem_rsid_bd}
\end{equation}
\end{lem}
\begin{IEEEproof}
Fix $\epsilon>0$. Let $m>\max\{\gamma^{2}(m_{0}-2),\,0\}$, and $\bar{F}\in\mathcal{P}_{+}$.
Let
\begin{align*}
\lambda & :=m\left(1-\frac{1}{\gamma}\right),\\
\zeta & :=\frac{\gamma-1}{2\gamma-1},\\
\tilde{n} & :=\left\lceil \frac{m}{\gamma^{2}}\right\rceil +1,\\
g_{0} & :=\bar{F}^{-1}(\zeta).
\end{align*}
Let $U\sim\mathrm{Unif}[\zeta,1]$, and 
\[
Z:=\bar{F}^{-1}(U)-g_{0}.
\]
Since $m>\gamma^{2}(m_{0}-2)$, we have $\tilde{n}\ge m_{0}$. By
\eqref{eq:lem_hyp}, let $\tilde{Z}\ge0$ be a random variable with
$F_{\tilde{Z}}\in\mathcal{P}_{+}^{*\infty}$, $F_{\tilde{Z}}\le F_{Z}^{*\tilde{n}}$,
and $\mathbf{E}[\tilde{Z}]\le\gamma\tilde{n}\mathbf{E}[Z]+\epsilon$.
Let $Z_{1},Z_{2},\ldots$ be i.i.d. copies of $Z$ independent of
$\tilde{Z}$. Let $N\sim\mathrm{Poi}(\lambda)$ independent of $\{Z_{i}\},\tilde{Z}$.
Let
\[
V:=mg_{0}+\tilde{Z}+\sum_{i=1}^{N}Z_{i}.
\]
We have $F_{V}\in\mathcal{P}_{+}^{*\infty}$.

We now show that $F_{V}\le\bar{F}^{*m}$. Let $W=Z$ with probability
$1-\zeta$, and $W=0$ with probability $\zeta$. Since $F_{g_{0}+W}\le\bar{F}$,
we have $F_{mg_{0}+\sum_{i=1}^{m}W_{i}}\le\bar{F}^{*m}$, where $\{W_{i}\}$
are i.i.d. copies of $W$. Let $M\sim\mathrm{Bin}(m,\,1-\zeta)$ independent
of $\{Z_{i}\}$. Then $\sum_{i=1}^{M}Z_{i}\stackrel{d}{=}\sum_{i=1}^{m}W_{i}$.
It will be shown in Appendix \ref{subsec:pf_sid_gap_2} that $N+\tilde{n}$
stochastically dominates $M$ (i.e., $F_{N+\tilde{n}}\le F_{M}$),
and hence 
\begin{align*}
\bar{F}^{*m} & \ge F_{mg_{0}+\sum_{i=1}^{m}W_{i}}\\
 & =F_{mg_{0}+\sum_{i=1}^{M}Z_{i}}\\
 & \stackrel{(a)}{\ge}F_{mg_{0}+\sum_{i=1}^{N+\tilde{n}}Z_{i}}\\
 & \ge F_{mg_{0}+\tilde{Z}+\sum_{i=1}^{N}Z_{i}}\\
 & =F_{V},
\end{align*}
where (a) is because we can couple $M$ together with $N$ such that
$N+\tilde{n}\ge M$.

We then bound $\mathbf{E}[V]=E(F_{V})$. We have
\begin{align*}
\mathbf{E}[V] & =mg_{0}+\mathbf{E}[N]\mathbf{E}[Z]+\mathbf{E}[\tilde{Z}]\\
 & \le mg_{0}+\left(\lambda+\gamma\tilde{n}\right)\mathbf{E}[Z]+\epsilon\\
 & =mg_{0}+\left(m\left(1-\frac{1}{\gamma}\right)+\gamma\left(\left\lceil \frac{m}{\gamma^{2}}\right\rceil +1\right)\right)\mathbf{E}[Z]+\epsilon\\
 & \le mg_{0}+\left(m+2\gamma\right)\mathbf{E}[Z]+\epsilon\\
 & =mg_{0}+\left(m+2\gamma\right)\frac{1}{1-\zeta}\int_{\zeta}^{1}\left(\bar{F}^{-1}(u)-g_{0}\right)du+\epsilon\\
 & =-2\gamma g_{0}+\left(m+2\gamma\right)\frac{1}{1-\zeta}\int_{\zeta}^{1}\bar{F}^{-1}(u)du+\epsilon\\
 & \le\left(m+2\gamma\right)\frac{1}{1-\frac{\gamma-1}{2\gamma-1}}E(\bar{F})+\epsilon\\
 & =\left(1+2\gamma m^{-1}\right)\left(2-\gamma^{-1}\right)E(\bar{F}^{*m})+\epsilon.
\end{align*}
The result follows from letting $\epsilon\to0$.
\end{IEEEproof}
\medskip{}

We now prove Theorem \eqref{thm:sid_gap}.
\begin{IEEEproof}
Fix $1<\alpha<2$ and let
\begin{equation}
\beta:=\frac{2(e(e-1)^{-1})^{\alpha}}{\alpha(\alpha-1)}+(e-1)^{-2}.\label{eq:beta_def}
\end{equation}
For $k\ge0$, let 
\[
\gamma_{k}:=\left(\frac{k+e}{k+e-1}\right)^{\alpha},
\]
\[
m_{k}:=\left\lfloor \beta\left(k+e-1\right)^{2\alpha}\right\rfloor .
\]
Note that
\begin{align*}
m_{k} & >\beta\left(k+e-1\right)^{2\alpha}-1\\
 & \ge\left(\frac{2(e(e-1)^{-1})^{\alpha}}{\alpha(\alpha-1)}+(e-1)^{-2}\right)\left(e-1\right)^{2\alpha}-1\\
 & >\frac{2(e(e-1)^{-1})^{\alpha}}{\alpha(\alpha-1)}\left(e-1\right)^{2\alpha}\\
 & =\frac{2e^{\alpha}(e-1)^{\alpha}}{\alpha(\alpha-1)}\\
 & \ge0,
\end{align*}
and hence $m_{k}\ge1$. We will prove inductively that for $k\ge0$,
\begin{equation}
\sup_{F\in\bigcup_{m\ge m_{k}}\mathcal{P}_{+}^{*m}}r_{\mathrm{ID}}(F)\le\gamma_{k}.\label{eq:th_ind}
\end{equation}
When $k=0$, $\gamma_{0}=(e(e-1)^{-1})^{\alpha}\ge e(e-1)^{-1}$,
and hence \eqref{eq:th_ind} holds by Theorem \ref{thm:infdiv_gap}.
Assume \eqref{eq:th_ind} holds for $k$. Fix any $m\ge m_{k+1}$.
We have
\begin{align*}
m & \ge\left\lfloor \beta\left(k+e\right)^{2\alpha}\right\rfloor \\
 & \ge\beta\left(k+e\right)^{2\alpha}-1\\
 & \ge\left(\frac{k+e}{k+e-1}\right)^{2\alpha}m_{k}-1\\
 & >\gamma_{k}^{2}(m_{k}-2).
\end{align*}
By Lemma \ref{lem:sid_gap_rec},
\begin{align*}
 & \sup_{F\in\mathcal{P}_{+}^{*m}}r_{\mathrm{ID}}(F)\\
 & \le\left(1+\frac{2\gamma_{k}}{m_{k}}\right)\left(2-\frac{1}{\gamma_{k}}\right)\\
 & \le\left(1+\frac{2(e(e-1)^{-1})^{\alpha}}{\beta\left(k+e-1\right)^{2\alpha}-1}\right)\left(2-\left(\frac{k+e-1}{k+e}\right)^{\alpha}\right)\\
 & \le\left(1+\frac{2(e(e-1)^{-1})^{\alpha}}{\beta\left(k+e-1\right)^{2}-1}\right)\frac{2(k+e)^{\alpha}-(k+e-1)^{\alpha}}{(k+e)^{\alpha}}\\
 & =\gamma_{k+1}\left(1+\frac{2(e(e-1)^{-1})^{\alpha}}{\beta\left(k+e-1\right)^{2}-1}\right)\left(1-\frac{(k+e+1)^{\alpha}-2(k+e)^{\alpha}+(k+e-1)^{\alpha}}{(k+e+1)^{\alpha}}\right)\\
 & \stackrel{(a)}{\le}\gamma_{k+1}\left(1+\frac{2(e(e-1)^{-1})^{\alpha}}{\beta\left(k+e-1\right)^{2}-1}\right)\left(1-\frac{\alpha(\alpha-1)(k+e+1)^{\alpha-2}}{(k+e+1)^{\alpha}}\right)\\
 & =\gamma_{k+1}\left(1+\frac{2(e(e-1)^{-1})^{\alpha}}{\beta\left(k+e-1\right)^{2}-1}\right)\left(1-\frac{\alpha(\alpha-1)}{(k+e+1)^{2}}\right)\\
 & \le\gamma_{k+1}\left(1+\frac{2(e(e-1)^{-1})^{\alpha}}{\beta\left(k+e-1\right)^{2}-1}-\frac{\alpha(\alpha-1)}{(k+e+1)^{2}}\right)\\
 & \stackrel{(b)}{\le}\gamma_{k+1},
\end{align*}
where (a) is because
\begin{align*}
 & (k+e+1)^{\alpha}-2(k+e)^{\alpha}+(k+e-1)^{\alpha}\\
 & =\int_{k+e}^{k+e+1}\alpha t^{\alpha-1}dt-\int_{k+e-1}^{k+e}\alpha t^{\alpha-1}dt\\
 & =\alpha\int_{k+e-1/2}^{k+e+1/2}\left((t+1/2)^{\alpha-1}-(t-1/2)^{\alpha-1}\right)dt\\
 & =\alpha\int_{k+e-1/2}^{k+e+1/2}\int_{t-1/2}^{t+1/2}(\alpha-1)s^{\alpha-2}dsdt\\
 & \ge\alpha(\alpha-1)(k+e)^{\alpha-2}\\
 & \ge\alpha(\alpha-1)(k+e+1)^{\alpha-2}
\end{align*}
since $1<\alpha<2$ and $s\mapsto s^{\alpha-2}$ is convex, and (b)
is because
\begin{align*}
 & \frac{\alpha(\alpha-1)}{(k+e+1)^{2}}\cdot\frac{\beta\left(k+e-1\right)^{2}-1}{2(e(e-1)^{-1})^{\alpha}}\\
 & =\alpha(\alpha-1)\frac{\beta-\left(k+e-1\right)^{-2}}{2(e(e-1)^{-1})^{\alpha}}\\
 & \ge1
\end{align*}
by \eqref{eq:beta_def}. Therefore, \eqref{eq:th_ind} holds for all
$k\ge0$ by induction.

Fix any $m\ge m_{\min}:=600$. Let 
\[
\alpha=1+(\ln m)^{-1}\le1+(\ln m_{\min})^{-1},
\]
\[
k=\left\lfloor \left(\frac{m}{\psi\ln m}\right)^{\frac{1}{2+2/\ln m}}-e+1\right\rfloor ,
\]
where
\[
\psi:=2(e(e-1)^{-1})^{1+(\ln m_{\min})^{-1}}+(e-1)^{-2}(\ln m_{\min})^{-1}.
\]
It can be checked that $k\ge0$. We have
\begin{align*}
m_{k} & \le\beta\left(k+e-1\right)^{2\alpha}\\
 & =\left(\frac{2(e(e-1)^{-1})^{\alpha}}{\alpha(\alpha-1)}+(e-1)^{-2}\right)\left(k+e-1\right)^{2\alpha}\\
 & \le\left(\frac{2(e(e-1)^{-1})^{\alpha}}{\alpha-1}+\frac{(e-1)^{-2}(\ln m_{\min})^{-1}}{\alpha-1}\right)\left(k+e-1\right)^{2\alpha}\\
 & \le\frac{\psi}{\alpha-1}\left(k+e-1\right)^{2\alpha}\\
 & =\psi(\ln m)\left(k+e-1\right)^{2+2/\ln m}\\
 & \le m.
\end{align*}
By \eqref{eq:th_ind},
\begin{align*}
 & \sup_{F\in\mathcal{P}_{+}^{*m}}r_{\mathrm{ID}}(F)\\
 & \le\gamma_{k}\\
 & =\left(\frac{k+e}{k+e-1}\right)^{\alpha}\\
 & =\frac{(k+e-1)^{\alpha}+\int_{k+e-1}^{k+e}\alpha t^{\alpha-1}dt}{(k+e-1)^{\alpha}}\\
 & \le\frac{(k+e-1)^{\alpha}+\alpha(k+e-1/2)^{\alpha-1}}{(k+e-1)^{\alpha}}\\
 & =1+\left(\frac{k+e-1/2}{k+e-1}\right)^{\alpha}\frac{\alpha(k+e-1/2)^{\alpha-1}}{(k+e-1/2)^{\alpha}}\\
 & \le1+\alpha\left(\frac{e-1/2}{e-1}\right)^{\alpha}\frac{1}{k+e-1/2}\\
 & \le1+\alpha\left(\frac{e-1/2}{e-1}\right)^{\alpha}\left(\left(\frac{m}{\psi\ln m}\right)^{\frac{1}{2+2/\ln m}}-1/2\right)^{-1}\\
 & \le1+\alpha\left(\frac{e-1/2}{e-1}\right)^{\alpha}\left(\frac{1}{\sqrt{\psi\ln m}}m^{\frac{1}{2+2/\ln m}}-1/2\right)^{-1}\\
 & =1+\alpha\left(\frac{e-1/2}{e-1}\right)^{\alpha}\left(\frac{1}{\sqrt{\psi\ln m}}\exp\left(\frac{(\ln m)^{2}}{2\ln m+2}\right)-1/2\right)^{-1}\\
 & =1+\alpha\left(\frac{e-1/2}{e-1}\right)^{\alpha}\left(\frac{1}{\sqrt{\psi\ln m}}\exp\left(\frac{\ln m}{2}-\frac{\ln m}{2\ln m+2}\right)-1/2\right)^{-1}\\
 & \le1+\alpha\left(\frac{e-1/2}{e-1}\right)^{\alpha}\left(\frac{1}{\sqrt{\psi\ln m}}\exp\left(\frac{\ln m}{2}-\frac{1}{2}\right)-1/2\right)^{-1}\\
 & =1+\alpha\left(\frac{e-1/2}{e-1}\right)^{\alpha}\left(\frac{1}{\sqrt{e\psi}}\sqrt{\frac{m}{\ln m}}-1/2\right)^{-1}\\
 & \le1+\alpha\left(\frac{e-1/2}{e-1}\right)^{\alpha}\frac{\frac{1}{\sqrt{e\psi}}\sqrt{\frac{m_{\min}}{\ln m_{\min}}}}{\frac{1}{\sqrt{e\psi}}\sqrt{\frac{m_{\min}}{\ln m_{\min}}}-1/2}\left(\frac{1}{\sqrt{e\psi}}\sqrt{\frac{m}{\ln m}}\right)^{-1}\\
 & =1+\alpha\left(\frac{e-1/2}{e-1}\right)^{\alpha}\frac{\sqrt{\frac{m_{\min}}{\ln m_{\min}}}\cdot\sqrt{\ln2}}{\frac{1}{\sqrt{e\psi}}\sqrt{\frac{m_{\min}}{\ln m_{\min}}}-1/2}\sqrt{\frac{\log m}{m}}\\
 & \le1+4.70662\sqrt{\frac{\log m}{m}}
\end{align*}
The bound also holds when $2\le m\le599$ by Theorem \ref{thm:infdiv_gap},
since $1+4.70662\sqrt{(\log m)/m}\ge e/(e-1)$ in this range. The
result follows.
\end{IEEEproof}
\medskip{}

A slightly curious consequence of Theorem \ref{thm:sid_gap} is that
if $H(X)=100$, then Theorem \ref{thm:pin_gap} implies that we can
divide $X$ into two i.i.d. pieces $Z_{1},Z_{2}$ with $H(Z_{1})\le70$.
In comparison, by Theorem \ref{thm:sid_gap}, we can divide $X^{100000}$
($100000$ i.i.d. copies of $X$) into $200000$ i.i.d. pieces, each
with entropy $\le56$, smaller than $70$. This shows that it is
easier to divide $X^{100000}$ into $200000$ i.i.d. pieces, than
to divide $X$ into two i.i.d. pieces.

If we can eliminate the $(1+2\gamma/m)$ term in \eqref{eq:lem_rsid_bd}
(which comes from rounding errors in bounding the cdf of binomial
and Poisson distributions), then we can improve the bound in \eqref{eq:th_rsid_bd}
to $1+O(1/\sqrt{m})$. We conjecture that this is the correct scaling
of the gap.
\begin{conjecture}
There exists a universal constant $c$ such that for all $m\in\mathbb{N}$,
\[
\sup_{F\in\mathcal{P}_{+}^{*m}}r_{\mathrm{ID}}(F)\le1+\frac{c}{\sqrt{m}}.
\]
\end{conjecture}
\medskip{}

\medskip{}

\section{Independent Identically-distributed Component Analysis\label{sec:iica}}

In this section, we introduce a variant of independent component analysis,
called independent identically-distributed component analysis (IIDCA).
Suppose $Z_{1},\ldots,Z_{n}$ are i.i.d. following an unknown distribution
$p_{Z}$, and $f(z_{1},\ldots,z_{n})$ is an unknown function (which
may not be injective). We observe the distribution of $X=f(Z_{1},\ldots,Z_{n})$,
or an estimate of the distribution, e.g. by i.i.d. samples of $X$.
The goal is to learn the distribution $p_{Z}$ and the function $f$.
Since the labeling of the values of $Z_{i}$ is lost, we can only
learn the entries of the pmf of $Z_{i}$, but not the actual values
of $Z_{i}$.

Our assumption is that the amount of information lost by the function,
i.e., $H(Z_{1},\ldots,Z_{n})-H(X)$, should be small. This means that
$f$ is close to being injective. This assumption is suitable if the
space of $X$ is rich enough to contain all information in $Z_{1},\ldots,Z_{n}$.
For the image example mentioned in Section \ref{subsec:iica_intro},
the assumption is suitable if the space of transformed images is rich
enough, for example, the transformed image has a larger size or a
richer color space. Another reason for this assumption is that it
is necessary to infer any information about $p_{Z}$, since if too
much information is lost by $f$, then the distribution of $X$ can
be quite arbitrary and reveals little information about $p_{Z}$.

Since minimizing $H(Z_{1},\ldots,Z_{n})-H(X)$ is equivalent to minimizing
$H(Z_{1})$, this gives the following optimization problem:
\begin{align}
\text{minimize} & H(Z_{1})\nonumber \\
\text{subject to} & Z_{1},\ldots,Z_{n}\;\mathrm{i.i.d.},\nonumber \\
 & H(X|Z_{1},\ldots,Z_{n})=0.\label{eq:prog_h}
\end{align}
By Theorem \ref{thm:pin_gap}, we can always achieve $H(Z_{1})\le(e/(e-1))H(X)/n+2.43$.
Nevertheless, it is difficult to solve \eqref{eq:prog_h} directly.
Therefore, we would consider a relaxation by allowing an arbitrary
cdf $F$ in place of the information spectrum $F_{\iota_{Z}}$:
\begin{align}
\text{minimize} & E(F)\nonumber \\
\text{subject to} & F\in\mathcal{P}_{+},\,F^{*n}\le F_{X}.\label{eq:prog_f}
\end{align}
After finding $F$, we can apply the procedures in the proof of Theorem
\ref{thm:pin_gap} to convert it to the desired distribution $p_{Z}$.
The gap between the optimal values of \eqref{eq:prog_h} and \eqref{eq:prog_f}
is bounded by $2.43$.

Another relaxation is to only require $p_{Z_{1},\ldots,Z_{n}}$ to
be majorized by $p_{X}$ (see \eqref{eq:maj}):
\begin{align}
\text{minimize} & H(p_{Z})\nonumber \\
\text{subject to} & p_{Z}^{\times n}\preceq p_{X}.\label{eq:prog_maj}
\end{align}
After finding $p_{Z}$, we can apply the procedures in the proof of
Theorem \ref{thm:pin_gap} (i.e., taking $\tilde{Z}=(Z,B)$, where
$B$ is given in \eqref{eq:pbdef}) to convert it to a distribution
$p_{\tilde{Z}}$, where $p_{X}$ is an aggregation of $p_{\tilde{Z}_{1},\ldots,\tilde{Z}_{n}}$,
and hence we can assume $H(X|\tilde{Z}_{1},\ldots,\tilde{Z}_{n})=0$.
The gap between the optimal values of \eqref{eq:prog_maj} and \eqref{eq:prog_f}
is bounded by $1.43$.

Nevertheless, \eqref{eq:prog_maj} is a non-convex problem. We propose
the following greedy algorithm. Let the size of the support of $X$
be $l$. We construct a probability vector $q=(q_{1},\ldots,q_{l})$,
$q_{1}\ge\cdots\ge q_{l}$ in the following recursive manner: for
$i=1,\ldots,l$, take 
\begin{equation}
q_{i}:=\max\left\{ t\ge0:\,(q_{1},\ldots,q_{i-1},t)^{\times n}\preceq p_{X}\right\} ,\label{eq:q_recur}
\end{equation}
where ``$\times n$'' is the $n$-fold tensor product of the vector
(treated as a vector in $\mathbb{R}^{i^{n}}$), and ``$\preceq$''
follows the same definition as \eqref{eq:maj} except that we do not
require the left hand side to be a probability vector. The optimal
$t$ can be found by binary search. This procedure continues until
$i=l$ or $q_{i}=0$. Note that $q_{1}=(\max_{x}p_{X}(x))^{1/n}$,
and $q_{i}$ is non-increasing since $(q_{1},\ldots,q_{i-1},t)^{\times n}\preceq p_{X}$
is a more stringent condition for larger $i$. We have $\sum_{j=1}^{i}q_{j}\le1$
(since $(q_{1},\ldots,q_{i})^{\times n}\preceq p_{X}$). If $\sum_{j=1}^{i-1}q_{j}<1$,
then the $q_{i}$ given by \eqref{eq:q_recur} satisfies $q_{i}>0$,
and at least one more equality in the inequalities in the definition
of $(q_{1},\ldots,q_{i})^{\times n}\preceq p_{X}$ is satisfied (compared
to $(q_{1},\ldots,q_{i-1})^{\times n}\preceq p_{X}$). To show this,
let $k\ge0$ be the largest integer such that 
\[
\max_{A\subseteq\{1,\ldots,i-1\}^{n}:\,|A|\le k}\sum_{\{a_{j}\}_{j}\in A}\prod_{j=1}^{n}q_{a_{j}}=\max_{B:\,|B|\le k}p_{X}(B),
\]
i.e., the $k$-th inequality in \eqref{eq:maj} is an equality. Assume
$\sum_{j=1}^{i-1}q_{j}<1$ and consider the $q_{i}$ that attains
the maximum in \eqref{eq:q_recur}. There exists $k'$ such that 
\begin{equation}
\max_{\begin{array}{c}
A\subseteq\{1,\ldots,i\}^{n}:\,|A|\le k',\\
\exists\{a_{j}\}_{j}\in A,\,j'\in\{1,\ldots,n\}\,\mathrm{s.t.}\,a_{j'}=i
\end{array}}\,\sum_{\{a_{j}\}_{j}\in A}\prod_{j=1}^{n}q_{a_{j}}=\max_{B:\,|B|\le k'}p_{X}(B),\label{eq:maxseqq}
\end{equation}
i.e., at least one inequality involving $q_{i}$ is an equality, or
else we can further increase $q_{i}$. If $k'\le k$, then there exists
$\{a_{j}\}_{j}\in\{1,\ldots,i\}^{n}$ where $a_{j'}=i$ for some $j'$,
and $\prod_{j=1}^{n}q_{a_{j}}$ is greater than or equal to the $k$-th
largest entry of $(q_{1},\ldots,q_{i-1})^{\times n}$ (and hence must
be equal since $(q_{1},\ldots,q_{i})^{\times n}\preceq p_{X}$), and
thus \eqref{eq:maxseqq} also holds for $k'=k+1$. Therefore we can
assume $k'>k$, and at least one more equality in the inequalities
in the definition of $(q_{1},\ldots,q_{i})^{\times n}\preceq p_{X}$
is satisfied. Therefore, when the procedure terminates, either $q_{i}=0$
or $i=l$ (all the $l$ inequalities in $(q_{1},\ldots,q_{i})^{\times n}\preceq p_{X}$
are equalities), both implying $\sum_{j=1}^{i}q_{j}=1$. Hence, when
the procedure terminates, the vector $q$ is a probability vector,
and we can take $p_{Z}=q$.

\medskip{}

\medskip{}

\section{Acknowledgement}

The author acknowledges support from the Direct Grant for Research,
The Chinese University of Hong Kong.

\medskip{}
\medskip{}

\appendix

\subsection{Proof of Proposition \ref{prop:imaj_maj} \label{subsec:pf_imaj_maj}}

Without loss of generality, assume $X,Y\in\mathbb{N}$, $p_{X}(1)\ge p_{X}(2)\ge\cdots$
and $p_{Y}(1)\ge p_{Y}(2)\ge\cdots$. For the ``only if'' part,
assume $p_{X}\preceq p_{Y}$. Fix any $\gamma$ and let $k_{X}\in\mathbb{Z}_{\ge0}$
be the largest integer satisfying $\sum_{x=1}^{k_{X}}p_{X}(x)\le\gamma$.
Define $k_{Y}$ similarly. Since $p_{X}\preceq p_{Y}$, we have $k_{X}\ge k_{Y}$.
We have
\begin{align*}
 & \int_{-\infty}^{\infty}2^{t}d\min\{F_{\iota_{X}}(t),\,\gamma\}\\
 & =k_{X}+\frac{\gamma-\sum_{x=1}^{k_{X}}p_{X}(x)}{p_{X}(k_{X}+1)}\\
 & \le k_{X}+1.
\end{align*}
Hence, if $k_{X}>k_{Y}$, we have $\int_{0}^{\infty}2^{t}d\min\{F_{\iota_{X}}(t),\,\gamma\}\ge\int_{0}^{\infty}2^{t}d\min\{F_{\iota_{Y}}(t),\,\gamma\}$.
It is left to consider $k_{X}=k_{Y}$. In this case,
\begin{align*}
 & \int_{-\infty}^{\infty}2^{t}d\min\{F_{\iota_{X}}(t),\,\gamma\}\\
 & =k_{X}+\frac{\gamma-\sum_{x=1}^{k_{X}}p_{X}(x)}{\sum_{x=1}^{k_{X}+1}p_{X}(x)-\sum_{x=1}^{k_{X}}p_{X}(x)}\\
 & \ge k_{X}+\frac{\gamma-\sum_{x=1}^{k_{X}}p_{Y}(x)}{\sum_{x=1}^{k_{X}+1}p_{Y}(x)-\sum_{x=1}^{k_{X}}p_{Y}(x)}\\
 & =\int_{-\infty}^{\infty}2^{t}d\min\{F_{\iota_{Y}}(t),\,\gamma\}.
\end{align*}
For the ``if'' part, assume $\int_{-\infty}^{\infty}2^{t}d\min\{F_{\iota_{X}}(t),\,\gamma\}\ge\int_{-\infty}^{\infty}2^{t}d\min\{F_{\iota_{Y}}(t),\,\gamma\}$.
Fix any $k_{Y}$. Let $\gamma=\sum_{y=1}^{k_{Y}}p_{Y}(y)$, and let
$k_{X}$ be defined as in the ``only if'' part. We have
\begin{align*}
k_{Y} & =\int_{-\infty}^{\infty}2^{t}d\min\{F_{\iota_{Y}}(t),\,\gamma\}\\
 & \le\int_{-\infty}^{\infty}2^{t}d\min\{F_{\iota_{X}}(t),\,\gamma\}\\
 & =k_{X}+\frac{\gamma-\sum_{x=1}^{k_{X}}p_{X}(x)}{p_{X}(k_{X}+1)}\\
 & <k_{X}+1.
\end{align*}
Hence $k_{Y}\le k_{X}$, and $\sum_{x=1}^{k_{Y}}p_{X}(x)\le\sum_{x=1}^{k_{X}}p_{X}(x)\le\gamma=\sum_{y=1}^{k_{Y}}p_{Y}(y)$.

\subsection{Proof of Proposition \ref{prop:e_noninc} \label{subsec:pf_e_noninc}}

Write $G_{i}=G_{F_{i}}$, and $G'_{i}(\gamma)$ for the left derivative
of $G_{i}$. By \eqref{eq:F_inv_logGp},
\begin{align*}
 & \int_{0}^{\infty}tdF_{i}(t)\\
 & =\int_{0}^{1}F_{i}^{-1}(\gamma)d\gamma\\
 & =\int_{0}^{1}\log G_{i}'(\gamma)d\gamma\\
 & =\int_{1}^{\infty}\int_{0}^{1}\min\{G_{i}'(\gamma),\,\xi\}d\gamma\cdot\frac{\log e}{\xi^{2}}d\xi.
\end{align*}
 Therefore, to prove $\int_{0}^{\infty}tdF_{1}(t)\ge\int_{0}^{\infty}tdF_{2}(t)$,
it suffices to prove $\int_{0}^{1}\min\{G_{1}'(\gamma),\xi\}d\gamma\ge\int_{0}^{1}\min\{G_{2}'(\gamma),\xi\}d\gamma$
for any $\xi\ge1$. Fix any $\xi\ge1$, and let $\eta:=\sup\{\gamma\in[0,1]:G_{1}'(\gamma)\le\xi\}$.
We have
\begin{align*}
 & \int_{0}^{1}\min\{G_{1}'(\gamma),\xi\}d\gamma\\
 & =\int_{0}^{\eta}G_{1}'(\gamma)d\gamma+(1-\eta)\xi\\
 & =G_{1}(\eta)+(1-\eta)\xi\\
 & \ge G_{2}(\eta)+(1-\eta)\xi\\
 & =\int_{0}^{1}\left(\mathbf{1}\{\gamma\le\eta\}G_{2}'(\gamma)+\mathbf{1}\{\gamma>\eta\}\xi\right)d\gamma\\
 & \ge\int_{0}^{1}\min\{G_{2}'(\gamma),\xi\}d\gamma.
\end{align*}

\subsection{Proof of Proposition \ref{prop:conv_imaj} \label{subsec:pf_conv_imaj}}

We use the following alternative definition of $G_{F}$:
\begin{align*}
G_{F}(\gamma) & =\int_{-\infty}^{\infty}2^{t}d\min\{F(t),\,\gamma\}\\
 & =\inf_{P_{Q|T}:\,Q\in[0,1],\,\mathbf{E}[Q]=\gamma}\mathbf{E}[2^{T}Q],
\end{align*}
where $T\sim F$ ($T$ has cdf $F$), and the supremum is over random
variables $Q\in[0,1]$ (which can be dependent of $T$) with $\mathbf{E}[Q]=\gamma$.
It is straightforward to check that the two definitions are equivalent.

To prove Proposition \ref{prop:conv_imaj}, it suffices to prove that
if $F_{1}\stackrel{\iota}{\preceq}\tilde{F}_{1}$, then $F_{1}*F_{2}\stackrel{\iota}{\preceq}\tilde{F}_{1}*F_{2}$
and $(1-\lambda)F_{1}+\lambda F_{2}\stackrel{\iota}{\preceq}(1-\lambda)\tilde{F}_{1}+\lambda F_{2}$.
Fix any $\gamma\ge0$. Let $T_{1}\sim F_{1}$, $T_{2}\sim F_{2}$,
$\tilde{T}_{1}\sim\tilde{F}_{1}$ mutually independent (hence $T_{1}+T_{2}\sim F_{1}*F_{2}$).
Fix any $\epsilon>0$, $\gamma\ge0$ and any random variable $Q\in[0,1]$
with $\mathbf{E}[Q]=\gamma$. We have
\begin{align*}
 & \mathbf{E}[2^{T_{1}+T_{2}}Q]\\
 & =\mathbf{E}\left[2^{T_{2}}\mathbf{E}[2^{T_{1}}Q\,|\,T_{2}]\right]\\
 & \stackrel{(a)}{\ge}\mathbf{E}\left[2^{T_{2}}G_{F_{1}}(\mathbf{E}[Q\,|\,T_{2}])\right]\\
 & \stackrel{(b)}{\ge}\mathbf{E}\left[2^{T_{2}}G_{\tilde{F}_{1}}(\mathbf{E}[Q\,|\,T_{2}])\right]\\
 & \stackrel{(c)}{\ge}\mathbf{E}\left[2^{T_{2}}(1-\epsilon)\mathbf{E}[2^{\tilde{T}_{1}}\tilde{Q}\,|\,T_{2}]\right]\\
 & =(1-\epsilon)\mathbf{E}[2^{\tilde{T}_{1}+T_{2}}\tilde{Q}]\\
 & \ge(1-\epsilon)G_{\tilde{F}_{1}*F_{2}}(\gamma),
\end{align*}
where (a) is by the alternative definition of $G_{F_{1}}$, and (b)
is by $F_{1}\stackrel{\iota}{\preceq}\tilde{F}_{1}$. For (c), we
let $\tilde{Q}\in[0,1]$ be a random variable such that $\mathbf{E}[\tilde{Q}\,|\,T_{2}=t_{2}]=\mathbf{E}[Q\,|\,T_{2}=t_{2}]$
and $\mathbf{E}[2^{\tilde{T}_{1}}\tilde{Q}\,|\,T_{2}=t_{2}]\le(1-\epsilon)^{-1}G_{\tilde{F}_{1}}(\mathbf{E}[Q\,|\,T_{2}=t_{2}])$
for any $t_{2}$ (this is possible due to the alternative definition
of $G_{\tilde{F}_{1}}(\mathbf{E}[Q\,|\,T_{2}=t_{2}])$). Also note
that $\mathbf{E}[\tilde{Q}]=\mathbf{E}[\mathbf{E}[\tilde{Q}\,|\,T_{2}]]=\mathbf{E}[Q]=\gamma$.
Hence,
\begin{align*}
G_{F_{1}*F_{2}}(\gamma) & =\inf_{Q\in[0,1],\,\mathbf{E}[Q]=\gamma}\mathbf{E}[2^{T_{1}+T_{2}}Q]\\
 & \ge(1-\epsilon)G_{\tilde{F}_{1}*F_{2}}(\gamma).
\end{align*}
Letting $\epsilon\to0$, we have $F_{1}*F_{2}\stackrel{\iota}{\preceq}\tilde{F}_{1}*F_{2}$.

To prove $(1-\lambda)F_{1}+\lambda F_{2}\stackrel{\iota}{\preceq}(1-\lambda)\tilde{F}_{1}+\lambda F_{2}$,
let $T:=T_{A}$ and $\tilde{T}=\mathbf{1}\{A=1\}\tilde{T}_{1}+\mathbf{1}\{A=2\}T_{2}$,
where $A=1$ with probability $1-\lambda$, $A=2$ with probability
$\lambda$. We have
\begin{align*}
 & \mathbf{E}[2^{T}Q]\\
 & =(1-\lambda)\mathbf{E}[2^{T_{1}}Q\,|\,A=1]+\lambda\mathbf{E}[2^{T_{2}}Q\,|\,A=2]\\
 & \ge(1-\lambda)G_{F_{1}}(\mathbf{E}[Q\,|\,A=1])+\lambda\mathbf{E}[2^{T_{2}}Q\,|\,A=2]\\
 & \ge(1-\lambda)G_{\tilde{F}_{1}}(\mathbf{E}[Q\,|\,A=1])+\lambda\mathbf{E}[2^{T_{2}}Q\,|\,A=2]\\
 & \stackrel{(a)}{\ge}(1-\lambda)(1-\epsilon)\mathbf{E}[2^{\tilde{T}}\tilde{Q}\,|\,A=1]+\lambda\mathbf{E}[2^{T_{2}}Q\,|\,A=2]\\
 & \ge(1-\epsilon)\mathbf{E}[2^{\tilde{T}}\tilde{Q}]\\
 & \ge(1-\epsilon)G_{(1-\lambda)\tilde{F}_{1}+\lambda F_{2}},
\end{align*}
where in (a), we let $\tilde{Q}\in[0,1]$ be a random variable such
that $\mathbf{E}[\tilde{Q}\,|\,A=1]=\mathbf{E}[Q\,|\,A=1]$ and $\mathbf{E}[2^{\tilde{T}}\tilde{Q}\,|\,A=1]\le(1-\epsilon)^{-1}G_{\tilde{F}_{1}}(\mathbf{E}[Q\,|\,A=1])$,
and $\tilde{Q}=Q$ if $A=2$. Hence
\begin{align*}
G_{(1-\lambda)F_{1}+\lambda F_{2}} & =\inf_{Q\in[0,1],\,\mathbf{E}[Q]=\gamma}\mathbf{E}[2^{T}Q]\\
 & \ge(1-\epsilon)G_{(1-\lambda)\tilde{F}_{1}+\lambda F_{2}}.
\end{align*}
The result follows from letting $\epsilon\to0$.

\subsection{Proof of the Claim in the Proof of Lemma \ref{lem:sid_gap_rec} \label{subsec:pf_sid_gap_2}}

We prove the following claim:
\begin{claim}
Let $\gamma>1$,
\[
\lambda:=n\left(1-\frac{1}{\gamma}\right),\,p:=\frac{\gamma}{2\gamma-1}.
\]
Let $a\in\mathbb{Z}_{\ge0}$ such that $a\ge n\gamma^{-2}$. Let $M\sim\mathrm{Bin}(n,p)$,
$N\sim\mathrm{Poi}(\lambda)$. Then we have $N+a+1$ stochastically
dominates $M$ (i.e., $F_{N+a+1}(t)\le F_{M}(t)$ for all $t$).
\end{claim}
\begin{IEEEproof}
We first check that $\mathbf{E}[N]+a=\lambda+a\ge np=\mathbf{E}[M]$.
We have
\begin{align*}
 & 1-\gamma^{-1}+\gamma^{-2}\\
 & =\frac{1-\gamma+\gamma^{2}}{\gamma^{2}}\\
 & \ge\left(\frac{\gamma}{2\gamma-1}\right)\left(\frac{(2\gamma-1)(1-\gamma+\gamma^{2})}{\gamma^{3}}\right)\\
 & =p\left(\frac{2\gamma^{3}-3\gamma^{2}+3\gamma-1}{\gamma^{3}}\right)\\
 & =p\left(1+\frac{(\gamma-1)^{3}}{\gamma^{3}}\right)\\
 & \ge p.
\end{align*}
Hence,
\begin{align*}
 & \lambda+a\\
 & \ge n\left(1-\frac{1}{\gamma}\right)+\frac{n}{\gamma^{2}}\\
 & =n\left(1-\gamma^{-1}+\gamma^{-2}\right)\\
 & \ge np.
\end{align*}
We use the following bound in \cite{zubkov2013complete}:
\[
F_{M}(k)\ge\Phi\left(\sqrt{2n}g_{\mathrm{B}}\left(\frac{k}{n}\right)\right)
\]
for $k=1,2,\ldots,n-1$, where $\Phi$ is the cdf of the standard
Gaussian distribution, and
\[
g_{\mathrm{B}}\left(r\right):=\mathrm{sign}(r-p)\sqrt{r\ln\frac{r}{p}+(1-r)\ln\frac{1-r}{1-p}}.
\]
We also use the following bound in \cite{short2013improved}:
\[
F_{N}(k)\le\Phi\left(\sqrt{2}g_{\mathrm{P}}\left(k+1\right)\right)
\]
for $k=0,1,2,\ldots$ (although \cite{short2013improved} requires
$k\ge1$, it is straightforward to check that the proof in \cite{short2013improved}
also works for $k=0$), where
\[
g_{\mathrm{P}}\left(t\right):=\mathrm{sign}(t-\lambda)\sqrt{\lambda-t+t\ln\frac{t}{\lambda}}.
\]
Define
\begin{align*}
g(t) & :=\mathrm{sign}(t-p)\sqrt{t\ln\frac{t}{p}+(1-t)\ln\frac{1-t}{1-p}}-\mathrm{sign}\left(t-1+\gamma^{-1}-\gamma^{-2}\right)\sqrt{1-\gamma^{-1}+\left(t-\gamma^{-2}\right)\ln\frac{t-\gamma^{-2}}{e\left(1-\gamma^{-1}\right)}}\\
 & \le\mathrm{sign}(t-p)\sqrt{t\ln\frac{t}{p}+(1-t)\ln\frac{1-t}{1-p}}-\mathrm{sign}\left(t-\frac{\lambda}{n}-\frac{a}{n}\right)\sqrt{\frac{\lambda}{n}+\left(t-\frac{a}{n}\right)\ln\frac{\left(t-\frac{a}{n}\right)}{e\frac{\lambda}{n}}}\\
 & =g_{\mathrm{B}}\left(t\right)-\frac{1}{\sqrt{n}}g_{\mathrm{P}}\left(nt-a\right).
\end{align*}
We now check that $F_{N+a+1}(k)\le F_{M}(k)$ for $k=0,1,2,\ldots$.
This is obvious for $k\le a$ since $F_{N+a+1}(0)=0$, and also obvious
for $k\ge n$ since $F_{M}(k)=1$. Hence, to check $F_{N+a+1}(k)\le F_{M}(k)$,
it suffices to check that
\[
g_{\mathrm{P}}\left(k-a\right)\le\sqrt{n}g_{\mathrm{B}}\left(\frac{k}{n}\right),
\]
for $a+1\le k\le n-1$. Therefore, it suffices to check that $g(t)\ge0$
for $\gamma^{-2}<t<1$. We consider 3 cases:
\begin{casenv}
\item $\gamma^{-2}<t\le p$: We have
\[
g(t)=\sqrt{1-\gamma^{-1}+\left(t-\gamma^{-2}\right)\ln\frac{t-\gamma^{-2}}{e\left(1-\gamma^{-1}\right)}}-\sqrt{t\ln\frac{t}{p}+(1-t)\ln\frac{1-t}{1-p}}.
\]
Let
\begin{align*}
\tilde{g}(t) & :=1-\gamma^{-1}+\left(t-\gamma^{-2}\right)\ln\frac{t-\gamma^{-2}}{e\left(1-\gamma^{-1}\right)}-t\ln\frac{t}{p}-(1-t)\ln\frac{1-t}{1-p}.
\end{align*}
We have
\begin{align*}
\frac{d\tilde{g}(t)}{dt} & =\ln\frac{t-\gamma^{-2}}{1-\gamma^{-1}}-\ln\frac{t}{p}+\ln\frac{1-t}{1-p}\\
 & =\ln\left(\frac{(t-\gamma^{-2})p(1-t)}{(1-\gamma^{-1})t(1-p)}\right)\\
 & =\ln\left(\left(-t+(\gamma^{-2}+1)-\gamma^{-2}t^{-1}\right)\frac{p}{(1-\gamma^{-1})(1-p)}\right)\\
 & =\ln\frac{-t+(\gamma^{-2}+1)-\gamma^{-2}t^{-1}}{(1-\gamma^{-1})^{2}}\\
 & =\ln\left(1-\frac{t-2\gamma^{-1}+\gamma^{-2}t^{-1}}{(1-\gamma^{-1})^{2}}\right)\\
 & =\ln\left(1-\frac{t^{-1}(t-\gamma^{-1})^{2}}{(1-\gamma^{-1})^{2}}\right)\\
 & \le0.
\end{align*}
Therefore $\tilde{g}(t)$ is non-increasing. It is clear that $g(p)\ge0$
(and hence $\tilde{g}(p)\ge0$). Hence, $\tilde{g}(t)\ge0$ and $g(t)\ge0$
for $\gamma^{-2}<t\le p$.\medskip{}
\item $p<t<1-\gamma^{-1}+\gamma^{-2}$: It is clear that both terms in $g(t)$
are non-negative.\medskip{}
\item $1-\gamma^{-1}+\gamma^{-2}\le t<1$: We have
\[
g(t)=\sqrt{t\ln\frac{t}{p}+(1-t)\ln\frac{1-t}{1-p}}-\sqrt{1-\gamma^{-1}+\left(t-\gamma^{-2}\right)\ln\frac{t-\gamma^{-2}}{e\left(1-\gamma^{-1}\right)}}.
\]
It is proved in Case 1 that $\tilde{g}(t)$ is non-increasing. It
is clear that $g(1-\gamma^{-1}+\gamma^{-2})\ge0$ (and hence $\tilde{g}(1-\gamma^{-1}+\gamma^{-2})\le0$).
Hence, $\tilde{g}(t)\le0$ and $g(t)\ge0$ for $1-\gamma^{-1}+\gamma^{-2}\le t<1$.
\end{casenv}
The result follows.
\end{IEEEproof}
\[
\]

\bibliographystyle{IEEEtran}
\bibliography{ref}

\end{document}